\documentclass[a4paper, useAMS, fleqn,usenatbib]{mnras} 

\usepackage[T1]{fontenc}
\usepackage{ae,aecompl}
\usepackage{graphicx}
\usepackage{float}
\usepackage[caption = false]{subfig}
\usepackage[usenames, dvipsnames]{color}
\usepackage{hhline}
\usepackage{multirow}
\usepackage{graphicx}
\usepackage{amsmath}	
\usepackage{gensymb} 
\usepackage{amssymb}	
\usepackage[normalem]{ulem}
\usepackage{capt-of}
\usepackage{times}
 
\newcommand{\massUnit}[1]{$ h^{-1} \; \rm{ M_{\odot}}$#1}
\newcommand{\distUnit}[1]{$ h^{-1} \; \rm {Mpc} $#1}
\title[Galaxy alignments with filament in EAGLE]{The Cosmic Ballet II: \enspace Spin alignment of galaxies and haloes with large-scale filaments in the EAGLE simulation}

\author[Punyakoti Ganeshaiah et al.]
{\parbox{\textwidth}{
    Punyakoti Ganeshaiah Veena,$^{1,2}$\thanks{E-mail:punyakoti.gv@gmail.com}
    Marius Cautun,$^{3}$ 
    Elmo Tempel,$^{2,4}$ \\
    Rien van de Weygaert$^{1}$
    and
    Carlos S. Frenk$^{3}$
\vspace{.1cm} }
\\ 
$^{1}$Kapteyn Astronomical Institute, University of Groningen,PO Box 800, 9747 AD, Groningen, The Netherlands\\
$^{2}$Tartu Observatory, University of Tartu, Observatooriumi 1, 61602 T$\tilde{o}$ravere, Estonia \\
$^{3}$Department of Physics, Institute for Computational Cosmology, University of Durham, South Road, Durham, DH1 3LE, UK\\
$^{4}$Leibniz-Institut f$\ddot{u}$r Astrophysik Potsdam (AIP), An der Sternwarte 16, 14482 Potsdam, Germany
}

\pubyear{2019}

\begin{document}
\label{firstpage}
\pagerange{\pageref{firstpage}--\pageref{lastpage}}
\maketitle

\begin{abstract}
We investigate the alignment of galaxies and haloes relative to cosmic web filaments using the EAGLE hydrodynamical simulation. We identify filaments by applying the NEXUS+ method to the mass distribution and the Bisous formalism to the galaxy distribution. Both web finders return similar filamentary structures that are well aligned and that contain comparable galaxy populations. EAGLE haloes have an identical spin alignment with filaments as their counterparts in dark matter only simulations: a complex mass dependent trend with low mass haloes spinning preferentially parallel to and high mass haloes spinning preferentially perpendicular to filaments. In contrast, galaxy spins do not show such a spin transition and have a propensity for perpendicular alignments at all masses, with the degree of alignment being largest for massive galaxies. This result is valid for both NEXUS+ and Bisous filaments. When splitting by morphology, we find that elliptical galaxies show a stronger orthogonal spin--filament alignment than spiral galaxies of similar mass. The same is true of their haloes, with the host haloes of elliptical galaxies having a larger degree of orthogonal alignment than the host haloes of spirals.  Due to the misalignment between galaxy shape and spin, galaxy minor axes are oriented differently with filaments than galaxy spins. We find that the galaxies whose minor axis is perpendicular to a filament are much better aligned with their host haloes. This suggests that many of the same physical processes determine both the galaxy--filament and the galaxy--halo alignments.
\end{abstract}


\section{Introduction}
The present study extends the analysis of \citet{GVeena2018} of the alignment of haloes with respect to cosmic web filaments. To this end, we explore whether the systematic alignment between the dark halo spins and their host filaments found in that study is preserved when studying the alignments of galaxy spins, and, in particular, we assess which factors may introduce differences in the spin--filament alignment of haloes and galaxies. 
Using the state-of-the-art EAGLE hydrodynamical simulation, we investigate
in parallel both the galaxy spin--filament and the halo spin--filament alignments as a function of galaxy mass and morphology.
Furthermore, to avoid an explicit dependence on the filament classifying method, we analyse the alignment of haloes and galaxies relative to filaments identified by two web finders: Nexus+ and Bisous \citep{cautun2013,TempelBisous2014}.

Galaxies in the Universe cluster together to form a web-like configuration known as the Cosmic Web. This large scale web is built up of dense superclusters connected by elongated filaments and sheet-like walls which surround underdense void regions \citep{Geller1983, Davis1985, shandarin1989, bond1996, vdw2008, Frenk2012, Liivamagi2012, TempelNature2014}. The cosmic web arises from the anisotropic gravitational collapse of primordial Gaussian density fluctuations, which evolve over billion of years into the highly complex and non-linear structures we observe today. 

The cosmic web is shaped by the gravitational tidal field, which determines the directions of anisotropic mass collapse. The same tidal field is also responsible for
spinning up haloes and galaxies. For example, during the linear phase of structure formation, the Tidal Torque Theory [TTT] \citep{hoyle1949, peebles1969, doroshkevich1970,white1984}, describes how the angular momentum of a protohalo is generated by the gravitational shear of the surrounding matter distribution. Specifically, the misalignment between the inertia tensor of the protogalaxy and the tidal tensor at that position generates a net spin (see \citealt{schaffer2009} for a review). 
Therefore, haloes and galaxies residing in different cosmic web environments acquire different spins.
Gradually, the angular momentum evolves until the time of turn around as the protohaloes and protogalaxies collapse and decouple from cosmic expansion. The spin thus acquired is mostly preserved even during the later stages of nonlinear evolution as the haloes develop into fully virialised entities.

The TTT and its extensions predict a direct correlation between the spin of haloes and the large-scale structure, such as an alignment of halo spin with the local directions of anisotropic collapse \citep[e.g.][]{efstathiou1979,barnes1987,Heavens1988,Lee2001,porciani2002,porciani2002TTT2,joneswey2009}. Using cosmological N-body simulations, \cite{aragon2007}, and shortly after \cite{hahn2007}, have confirmed that halo spins are preferentially aligned with the orientation of the cosmic filaments and walls in which they are located and they has been confirmed by numerous follow-up studies \citep[e.g.][]{codis2012,libeskind2013,trowland2013,romero2014,Wang2017a,Lee2018}. 

Of particular interest is the halo spin--filament alignment that shows a complex mass dependence, with high mass haloes having spins preferentially perpendicular to filaments while low mass haloes show the opposite trend, with their spins being preferentially parallel to filaments.
The halo mass at which this transition happens is known as the \emph{spin-flip transition mass}, or, in short, spin-flip mass. This transition mass increases with decreasing redshift \citep[e.g. see][]{codis2012,Wang2018a} and is ${\sim}1 \times 10^{12}$ \massUnit{} at present day, with the exact value differing by up to a factor of several between different studies. Furthermore, the spin-flip mass is highest for haloes in thick filaments and is up to an order of magnitude lower for haloes in thin and tenuous filaments \citep{GVeena2018}. The dichotomy in spin--filament alignment between low- and high-mass haloes has been attributed to various processes related to late-time accretion \citep[e.g.][]{Welker2014,codis2015,Laigle2015,Wang2017a,GVeena2018}. High-mass haloes form recently and accrete most of their mass along the filaments they reside in, which results in a net spin gain that is preferentially perpendicular on the filament axis. In contrast, low-mass haloes accumulated most of their mass at higher redshift when they might have been found in cosmic sheets and any present day mass accretion imparts a net spin along their host filament.

Extending the halo spin--filament alignment results to galaxies is not trivial since the spin of many galaxies is poorly aligned with that of their host halo \citep[e.g.][]{Velliscig2015,Tenneti2016,Shao2016,Chisari2017}.
As gas streams enter the inner regions of the halo, they gain most of their angular momentum through non-linear torques and dissipation and galaxy spin is affected by disc instabilities and feedback, such as gas outflows due to supernovae  \citep{Danovich2015}. 
Therefore, though galaxies and haloes were subjected to the same tidal fields that generated the initial angular momentum, we expect galaxy spins to deviate from the their host halo spins.        
\citet{hahn2010} found that in an AMR hydrodynamical simulation, massive discs have spins aligned along the filaments. 
\citet{codis2012, Dubois2014, Welker2014} study spin--filament alignment for galaxies between redshift 1.2 and 1.8 using the Horizon AGN simulation. They report a galaxy spin transition from parallel to perpendicular at a stellar mass of $\sim 3 \times10^{10} $\massUnit{} and find that the spin of 
blue galaxies is preferentially parallel to the nearest filament whereas the spun of red galaxies shows a preferential perpendicular alignment. \citet{Codis2018} find that the parallel alignment signal for low mass galaxies is weak and decreases with time whereas the strength of the orthogonal alignment of high mass galaxies increases with time. 
\cite{Wang2018} show that the spin of low mass, blue galaxies in the Illustris-1 hydrodynamical simulation are preferentially along the filament axis whereas the massive, red galaxies have spins preferentially perpendicular. 

This trend was also confirmed observationally by \citet{tempel2013}, who found that the spins of high 
mass ellipticals are preferentially 
perpendicular while those of bright spiral galaxies
are preferentially parallel to their host filaments \citep[see also][]{Cervantes-Sodi2010,jones2010,elmonoam2013,Zhang2013,Zhang2015, Pahwa2016, Hirv2017}. 

In this paper we address how secondary baryonic processes alter the spin of galaxies, initially imparted by tidal torques, and hence its alignment with the cosmic filaments in which the galaxies reside. Mainly, we address the following questions:  
\begin{itemize}
\item Do galaxies exhibit a mass dependent spin alignment in hydrodynamical simulations? 
\item How does the addition of baryons alter the transition mass of the halo spin-filament alignment?
\item Does the galaxy spin--filament alignment signal depend on the filament identification method?
\item If a galaxy orientation with respect to its parent filament is known, is it
possible to infer the orientation of its host halo?
\end{itemize}
       
In this study we carry out a detailed analysis of galaxy and halo spin--filament alignments in the EAGLE hydrodynamical simulation \citep{schaye2015,crain2015}. We employ tow methods to identify the filamentary pattern: NEXUS+, which uses the total matter density field, and Bisous, which uses the the galaxy distribution. It is essential to compare the two cosmic web tracers, because while the matter distribution generates the tidal field, observational surveys trace only the galaxy distribution, which is a sparse and biased tracer of the total matter distribution. Therefore, we compare the spin alignments with respect to filaments detected in both matter and galaxy distributions. Further, we investigate the correlation between galaxy spin--filament alignment and galaxy morphology, and, whenever possible, compare against observations. 

The structure of the paper is as follows.
In Section 2, we describe the EAGLE hydrodynamical simulation and the galaxy and halo samples used in our analysis, and give a short overview of the cosmic web extraction algorithms we employ. In Section 3 we compare the NEXUS and Bisous filament populations and their corresponding haloes and galaxies. Section 4 presents the main results on alignments of haloes and galaxies with the orientation of their host filament. Finally, in Section 5 we give a brief summary of our study and discuss its implications.

\begin{figure*}
    \centering
    \includegraphics[width=.98\columnwidth]{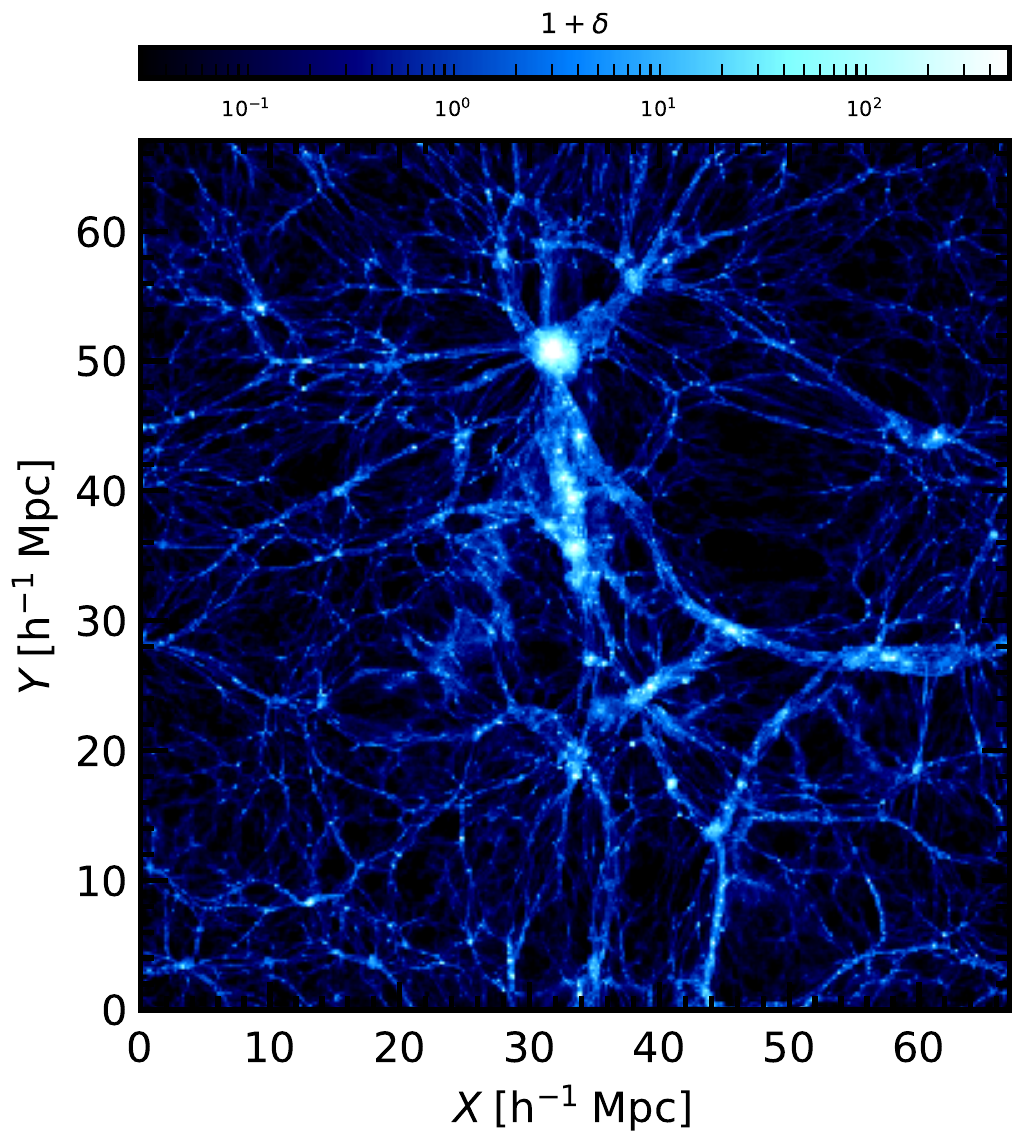}
	\includegraphics[width=.98\columnwidth]{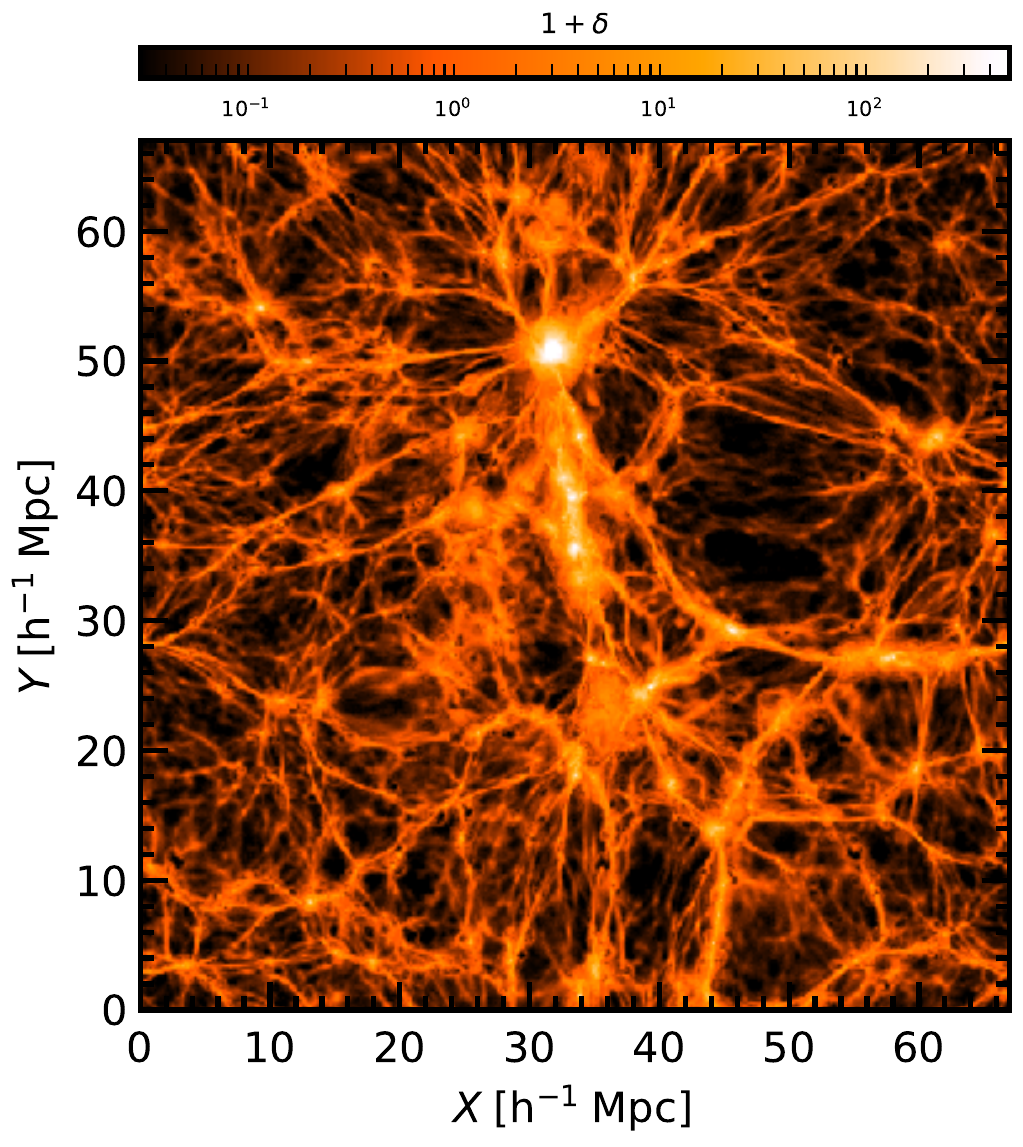} \\
    \includegraphics[width=.98\columnwidth]{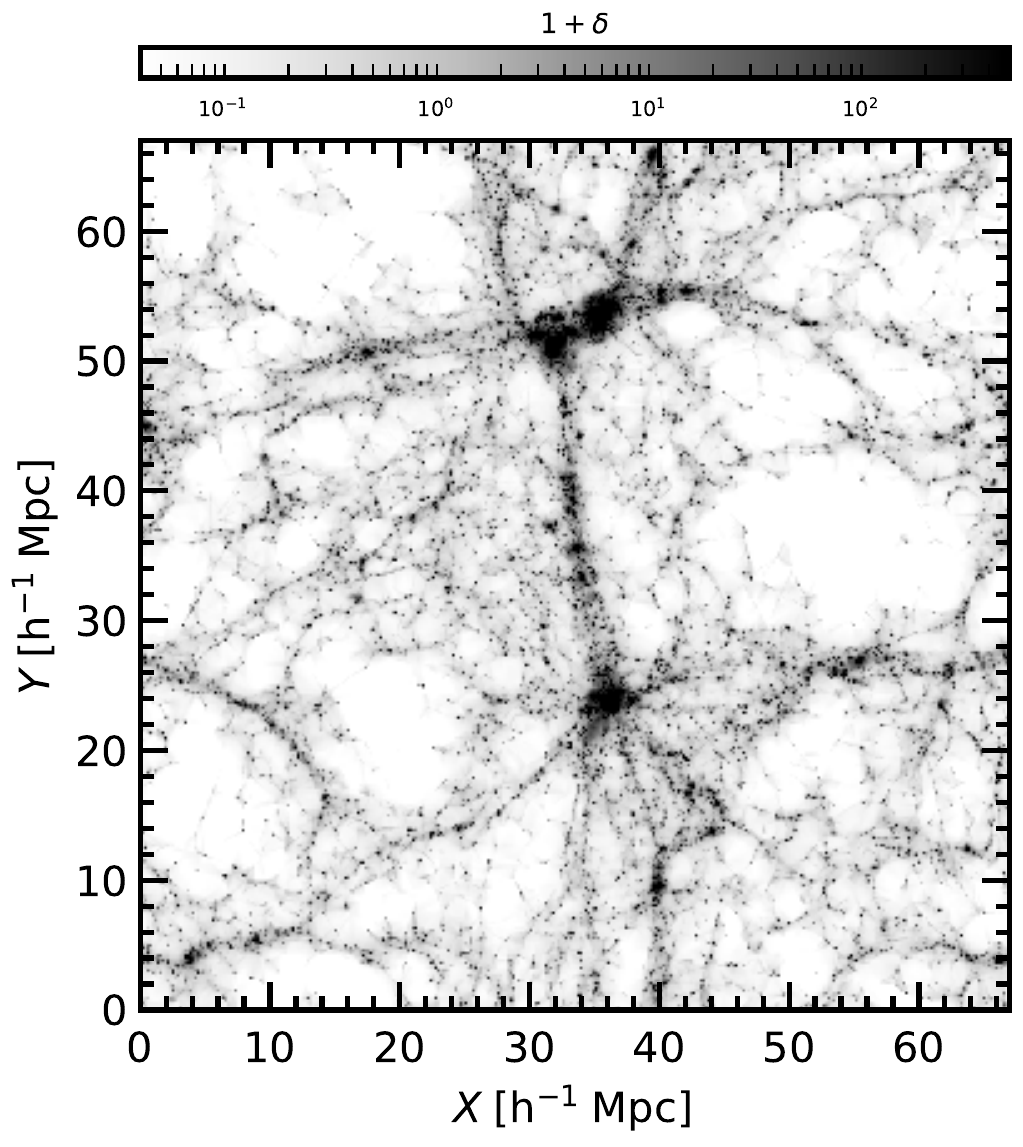}
    \caption{\textbf{Dark matter, gas and stellar density fields.}
    Dark matter (top-left) and gas (top-right) density distributions in a thin slice ($ 132 \; h^{-1} \; \rm {kpc}$) of the EAGLE simulation. \textit{Lower Panel:} Stellar density field in a thick slice ($10 \; h^{-1} \; \rm {Mpc})$ centred on the thin slice shown in the top panels. 
    The logarithmically scaled colour bar represents the density contrast, $1+\delta$.
    } 
    \label{fig:density_field}
    \vskip -.2cm
\end{figure*} 

\vspace{-.2cm}
\section{Data}
In this section, we first describe the EAGLE simulation and the procedure used to extract the galaxy and halo samples. Then, we give a short overview of the two web identification methods, NEXUS+ and Bisous.

\vspace{-.2cm}
\subsection{EAGLE  simulation}
\label{sec:Eagle_simulation}
Our analysis makes use of the largest box (Ref-L0100N1504) of the EAGLE cosmological simulation of galaxy formation. EAGLE follows the baryonic processes that shape galaxy formation and evolution and thus allows us to study the influence of the large scale tidal fields on the underlying physics of galaxy formation and galaxy properties such as spin, shape and morphology. 
The simulation follows the evolution of $1504^3$ dark matter particles and an initial equal number of gas particles in a periodic box of $67.7$ \distUnit{} side length, which is large enough to resolve a multitude of large scale environments. Each dark matter particle has a mass of 
$6.57 \times 10^{6}$ \massUnit{} and each gas particle has an initial mass of $1.2 \times 10^{6}$ \massUnit{}. 

The simulation is based on the $\Lambda$CDM cosmology and assumes the \citet{planck2015} cosmological parameters, which take the following values: $\Omega_{\Lambda} =0.693$, $\Omega_{\rm M} = 0.307$ and $\Omega_{\rm b} = 0.0455$, $\sigma_8=0.8288$ and $h= 0.6777$, where $H_0 = 100~h~\rm{ km \; s^{-1} Mpc^{-1}} $ is the Hubble's constant at present day.
The EAGLE project was run using a modified version of the GADGET code \citep{springel2005} and it includes numerous baryonic processes relevant for galaxy formation that have been calibrated to match: (a) the observed galaxy stellar mass function, (b) the distribution of galaxy sizes, and (c) the observed relation between galaxy and central black hole mass \citep[for details see][]{schaye2015,crain2015}.

A visualisation of the dark matter, gas and density fields in the EAGLE simulation can be seen in \autoref{fig:density_field}. This has been obtained by applying the Delaunay Tessellation Field Estimator software \citep{schaap2000, cautun2011} to the dark matter, gas and star particle distribution to interpolate their respective density fields to a regular grid. 
While dark matter and gas trace the same structures on very large scales, gas is more diffused compared to dark matter (DM), especially in the high density 
regions. This is due to processes such as supernovae and AGN feedback which heat up the gas and make it less dense. In contrast, dark matter is not directly affected by such processes and can therefore form denser and more compact structures. See \citet{haider2016} for a more detailed discussion of the effects of feedback on the general properties of large-scale structures.

The bottom panel of \autoref{fig:density_field} shows the stellar density field in a $10$ \distUnit{} thick slice. The stars are mostly found in the very centre of haloes and taking a thin slice through their distribution results in predominantly empty space. Thus, to appreciate the outline of the cosmic web, we show the galaxies in a much thicker slice than the one used to show the DM or gas distributions in the top row of \autoref{fig:density_field}. The stars are predominantly found in regions with high DM and gas densities, which is where haloes are mostly found.

\begin{figure}
    \centering
    \vskip -.3cm
    \includegraphics[width=\columnwidth]{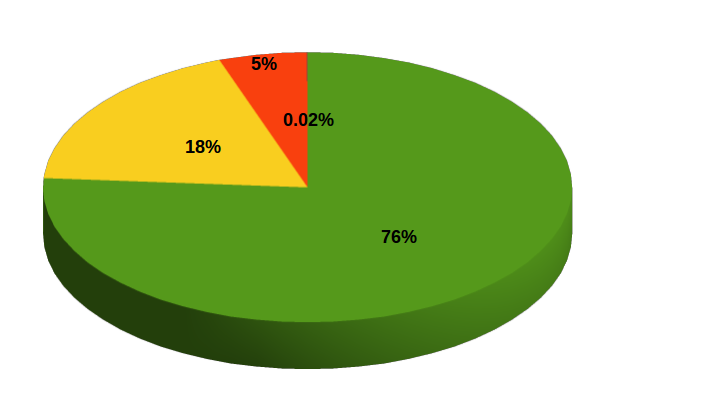}
    \vskip -0.5cm
    \includegraphics[width=\columnwidth]{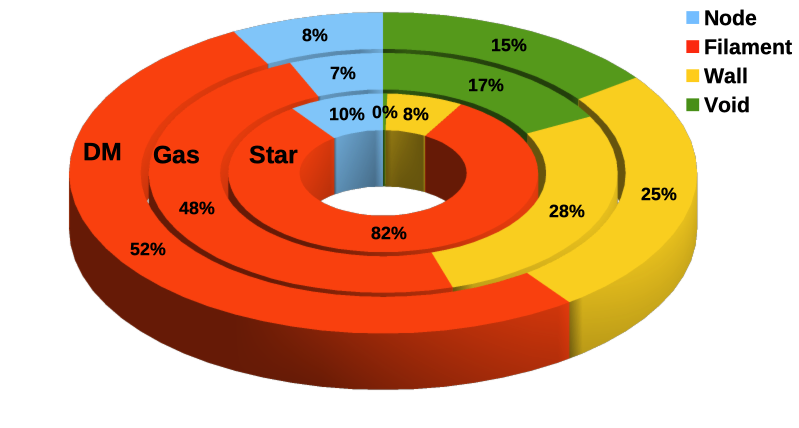}
    \vskip -0.5cm
    \caption{ \textbf{Volume and mass fractions of the cosmic web.} The results are for web environments identified by applying the NEXUS+ method to the DM density field. The top panel shows the volume fraction occupied by each web environment. The bottom panel shows the mass fraction of DM, gas and stars in each environment. The exact values are given in \autoref{table:mass_volume_table}. 
    }   
    \label{fig:vol_mass_fraction}
\end{figure}

\autoref{fig:vol_mass_fraction} compares the volume and mass fraction in different components of the cosmic web as identified by NEXUS+. We find that 76\% of the volume in the universe is occupied by voids followed by walls (18\%), filaments (6\%) and clusters (0.02\%) which is in good agreement with the \citet{cautun2014} results based on DM-only cosmological simulations. In terms of mass, filaments contain most of the mass distribution of the universe: around 50\% of the DM and gas, and 82\% of stars. 
The high mass fraction of stars is a consequence of the fact that most haloes more massive than a few $\times 10^{11}$ \massUnit{} are found in filaments.
We also notice that compared to DM mass fraction there is slightly less gas in nodes and filaments, for example filaments contain roughly 52\% of the DM and 47\% of the gas budget. While initially gas follows the DM distribution, winds and feedback processes during galaxy formation heat up, push and disperse the gas from nodes and filaments into adjacent walls and voids \citep{haider2016,Martizzi2018}.

In \autoref{table:mass_volume_table} we tabulate mass and volume fractions of DM, gas and stars in different cosmic web environments. The DM represents the vast majority of the cosmic mass budget, however it cannot be observed directly. To study the extent to which the gas distribution traces the same cosmic web as the DM, we applied the NEXUS+ method separately to the DM and gas density fields. In general, we find good agreement between the mass and volume fraction in the two web types indicating that the gas distribution is a good tracer on large scales of the total density. The only large difference is for nodes, where nodes identified in the gas distribution contain ${\sim}10\%$ less DM, gas and stars than nodes identified in the DM distribution. For the other web environments, the differences between the DM and gaseous cosmic web are much smaller.

\begin{table}
    \centering
    \caption{\textbf{Mass and volume fractions of the cosmic web.} The web environments were identified using the NEXUS+ method applied to the DM (first 4 rows) and to the gas (rows 5 to 8) distributions. The last row corresponds to Bisous filaments identified using the galaxy distribution.}
    \label{table:mass_volume_table}
\begin{tabular}{ @{} ccccc @{}} 
\hline
\hline
Environment & Volume [\%] & \multicolumn{3}{c}{Mass [\%]}  \\ [.05cm]
  &   & [DM]  & [Gas] & [Stars]    \\
\hline \\  [.05cm]
\multicolumn{5}{c}{\textbf {NEXUS+ applied to the DM density}}  \\
\hline
    {Node} &0.02 & 7.9 & {7.1}& {10}  \\
    {Filament}& 5.4   & 52 & { 47}& {82} \\
    {Wall} & 19   & 25 & {28 }& {8.0} \\ 
    {Void} & 76  & 15 & { 17}& {0.43} \\
\hline \\ [.05cm]
\multicolumn{5}{c}{\bf NEXUS+ applied to the gas density} \\  
\hline  
    {Node}& 0.02  & {7.2} & 6.5&  {8.7} \\
    {Filament } &  5.9 & {53 } & 48 & {82} \\
    {Wall} & 18 & {25 } & 28 &  {8.8}  \\ 
    {Void} & 76  &  {15}& 17& {0.43}  \\ 
\hline \\ [.05cm]
\multicolumn{5}{c}{\bf Bisous applied to the galaxy distribution} \\
\hline
Filament &  5.1 & 45 &   41 & 70 \\ 
\hline \\
\end{tabular}
\end{table}

\subsection{Filament population}
\label{subsec:fila_population}
To detect large scale filaments in the EAGLE  simulation, we use two different web identification algorithms: NEXUS+ \citep{cautun2013}  and Bisous \citep{TempelBisous2014, tempel2016}.
These algorithms detect the filamentary network based on two fundamentally different approaches. 
NEXUS+ is a geometric technique that detects filaments based on the morphology of the density field. Bisous is a statistical technique that extracts the filamentary network by applying a statistical model directly on the distribution of galaxies.

We wish to probe how the differences in these two filament populations influence the results on galaxy spin and shape alignments. Below we describe briefly the working and the implementation of the two formalisms. 

\subsubsection{ Filament detection using NEXUS+}
The MMF/NEXUS \citep{aragon2007MMF,cautun2013} technique uses the geometry of the matter distribution to identify the cosmic web environments. Among its most defining features, NEXUS uses the \textit{Scale-Space formalism} to identify web environments at several scales. The method has its roots in the field of medical imaging \citep[see e.g.][]{sato1998, Li2003} and has been adapted to astronomy by \citet{aragon2007MMF} under the name the Multiscale Morphology Filter (MMF). The variant that we use in this paper, the NEXUS+ method, is an advanced version of the MMF technique and has been developed to better account for the many orders of magnitude variation in the large-scale density field. 

The main advantage of the NEXUS+ formalism is that it simultaneously identifies cosmic web morphology at several spatial scales. Thus, it deals with the multiscale nature of the cosmic web, which is a consequence of hierarchical structure formation and which represents a crucial aspect of the connection between the cosmic web and halo/galaxy properties. The NEXUS+ method is based on using the eigenvalues and eigenvectors of the local Hessian matrix for a range of smoothing scales, which are then used to identify the web environments.

The steps involved in the NEXUS+ formalism are as follows (for more details see \citealt{cautun2013}):
\begin{enumerate}
    \item[\textbf{Step 1:}] Apply a Log-Gaussian filter of width $R_n$ to the cosmic density contrast field, $\delta=\tfrac{\rho}{\bar{\rho}}-1$, where $\rho$ and $\bar{\rho}$ denote the local and mean background density, respectively. The Log-Gaussian filter consists of calculating the density logarithm, $\log(1+\delta)$, smoothing the logarithm with a Gaussian filter of size $R_n$ and then calculating the smoothed density, $\delta_{R_n}$, from the smoothed density logarithm. 
    
    \item[\textbf{Step 2:}] Next, the Hessian matrix at each position, $H_{ij, R_n}(\mathbf{x})$, is calculated using: 
    \begin{eqnarray}
    H_{ij, R_n} &=& R_n^2 \frac{\partial^2 \delta_{R_n}(\mathbf{x})}{\partial x_i \partial x_j} 
    \end{eqnarray}
    Here, the re-normalization by $R_n^{2}$ ensures that the Hessian is weighted identically at different scales. In this paper, we implement filter scales in the range 0.5 to 4.0 \distUnit. We go from the smallest relevant scale to the upper limit of $4$\distUnit, which allows us to identify large filaments. 

    \item[\textbf{Step 3:}] A node, filament and wall characteristic is assigned at each point $\mathbf{x}$ based on the nature of the Hessian matrix eigenvalues, $\lambda_1 \le  \lambda_2 \le \lambda_3 $. These are used to define the web environment signature, $ S_{R_n}(\mathbf{x})$. The exact equation for defining environments is complex, but, qualitatively, nodes corresponds to regions with $\lambda_1 \approx \lambda_2 \approx \lambda_3 < 0 $, filaments to regions with $\lambda_1 \approx \lambda_2 < 0 $ and $\lambda_2 \ll \lambda_3$, and walls to $ \lambda_1  < 0 $ and $\lambda_1 \ll \lambda_2$. In particular, orientation of filaments corresponds to the eigenvector, $\mathbf{e_{n3}}$, along the slowest direction of collapse.

    \item[\textbf{Step 4:}] Subsequently, steps 1 to 3 are repeated for a set of scales $[R_0, R_1,...R_N ]$ and at each scale the environment signature, $ S_{R_n}(\mathbf{x}) $, is computed. 

    \item[\textbf{Step 5:}] The environmental signature for the various filter scales is combined together to obtain a scale independent signature, $S(\mathbf{x})$. This is defined as the maximum value of all the scales:
    \begin{eqnarray}
    \mathcal{S}(\mathbf{x}) = \max_{\rmn{levels\;}n} \mathcal{S}_{R_n}(\mathbf{x})\,.
    \label{eq:total_response}
    \end{eqnarray}

    \item[\textbf{Step 6:}] Finally, a threshold signature is used to determine the validity of an identified morphology. Signatures greater than the threshold are considered valid structures and rest are discarded.  
\end{enumerate}
From this method, we find a total of 6394 galaxies in NEXUS+ filaments which is $\sim 67\%$ of the total galaxy sample.

\subsubsection{Filament detection using Bisous}
The Bisous filament finding algorithm \citep{TempelBisous2014} works by randomly distributing a large number of fixed radius cylinders onto a galaxy distribution, and estimating how likely it is that each cylinder corresponds to a cosmic filament. This is achieved by comparing the number of galaxies inside the cylinder with the number just outside the cylinder, with filaments corresponding to a large galaxy density contrast inside a cylinder. The Bisous method
is based on a marked point process which was originally designed to extract spatial patterns \citep{stoica2005}. A marked point process is a point process with an additional parameter or a mark associated with every point. In the context of the Bisous formalism, centers of cylinders outlining the galaxy distribution are treated as points whose mark is related to the length, radius and orientation of the cylinder.
The cosmic web filamentary network is then constructed by selecting the most connected and well aligned cylinders.
Briefly, the following steps are involved in determining the Bisous filamentary network from the galaxy distribution: 

\begin{enumerate}
\item[\textbf{Step 1:}] Multiple Markov-Chain Monte-Carlo (MCMC) simulations are
performed to obtain the cylinder configurations that outline the filamentary network based on the distribution of galaxies. These cylinders, which eventually make up the filamentary network, have a fixed radius, varying length and orientation. The probability that a cylinder should be retained in the 
filamentary network is determined by the distribution of the galaxies within 
each cylinder and its connectivity and alignment to its neighbouring cylinders.

\item[\textbf{Step 2:}]  Using the MCMC simulations, a visit map is then determined, which gives the probability that a certain region or galaxy belongs to the filamentary network.

\item[\textbf{Step 3:}] The ridges of the visit map are considered as filament spines and a filamentary network for the given galaxy distribution is constructed\footnote{A visualisation of the steps 1 to 3 can be seen here:~\url{http://www.aai.ee/~elmo/sdss-filaments/sdss_filaments.mp4}}. Galaxies with high visit map values and those which are also placed within a certain fixed distance from the filament spine are identified as galaxies in Bisous filaments. 
\end{enumerate}
For a detailed explanation of the mathematical framework of the Bisous model, we refer to \citet{TempelBisous2014,tempel2016}.

In the current study, we apply the Bisous model to the spatial distribution of all EAGLE galaxies with stellar masses above $1 \times 10^8$ \massUnit{}. We define Bisous filament galaxies as all the galaxies that are within a distance of 1 Mpc from the filament spine using only locations with a visit map value larger than $0.05$. In total we find that there are 5988 such central galaxies, which is ${\sim}63\%$ of the total sample. The algorithm also computes the orientation of the filaments, denoted as $\mathbf{e_{b3}}$, as the unit vector along the filament spine. 

The Bisous methodology has been successfully applied to SDSS to look for galaxy-filament alignments \citep{elmonoam2013, TempelTamm2015} and satellite alignments \citep{TempelQuo2015}. Applying Bisous to a $\Lambda$~CDM hydrodynamical simulation represents the next step towards comparing the galaxy spin--filament alignment between theory and observations.

\subsection{Halo and galaxy populations}
\label{subsec:galaxy_population}
Haloes and galaxies are extracted from the EAGLE simulation  using the Friends-of-Friends and SUBFIND algorithms \citep{springel2001} as described in \citet{mcalpine2016}. Initially, DM clumps are identified using the FoF method by adopting a linking length of 0.2 times the average separation of DM particles. Every baryonic particle is then allotted to the FoF group to which its nearest DM particle belongs. The SUBFIND algorithm then identifies gravitationally bound substructures within these FOF groups. Therefore, every FoF group may have more than one substructure and the most gravitationally bound (least gravitational potential) substructure is labelled as the central galaxy and the rest are labelled as satellites. 

For our analysis we use only the central galaxies above a stellar mass of $5  \times 10^{8} \; h^{-1} M_{\odot}$ and their corresponding DM subhaloes (hereafter haloes). We choose this mass limit to ensure we have at least 300 stellar particles, enough to resolve the inner stellar and gas distributions of a galaxy and also to achieve convergence for properties such as angular momentum, shape and morphology  \citep[see e.g.][]{bett2007}. 

\subsubsection{Halo and galaxy masses}
The radius of a DM halo, $R_{200}$ is defined as the radius from the halo centre within which the average halo density is 200 times the
critical density of the universe. The mass of a halo, $M_{200}$, is calculated as the total mass inside the $R_{200}$ radius.
For the galaxies, in order to avoid baryonic particles that may belong to the intra cluster region, an aperture mass is computed. The stellar mass, $M_\mathrm{star}$, corresponds to the stellar mass within an aperture of $10$ kpc while the gas mass, $M_\mathrm{gas}$, corresponds to the gas mass within an aperture of $30$ kpc. We choose these definitions as they are similar to the observational measurements of stellar and gas disc components of galaxies.

\begin{figure}
    \includegraphics[width=\columnwidth]{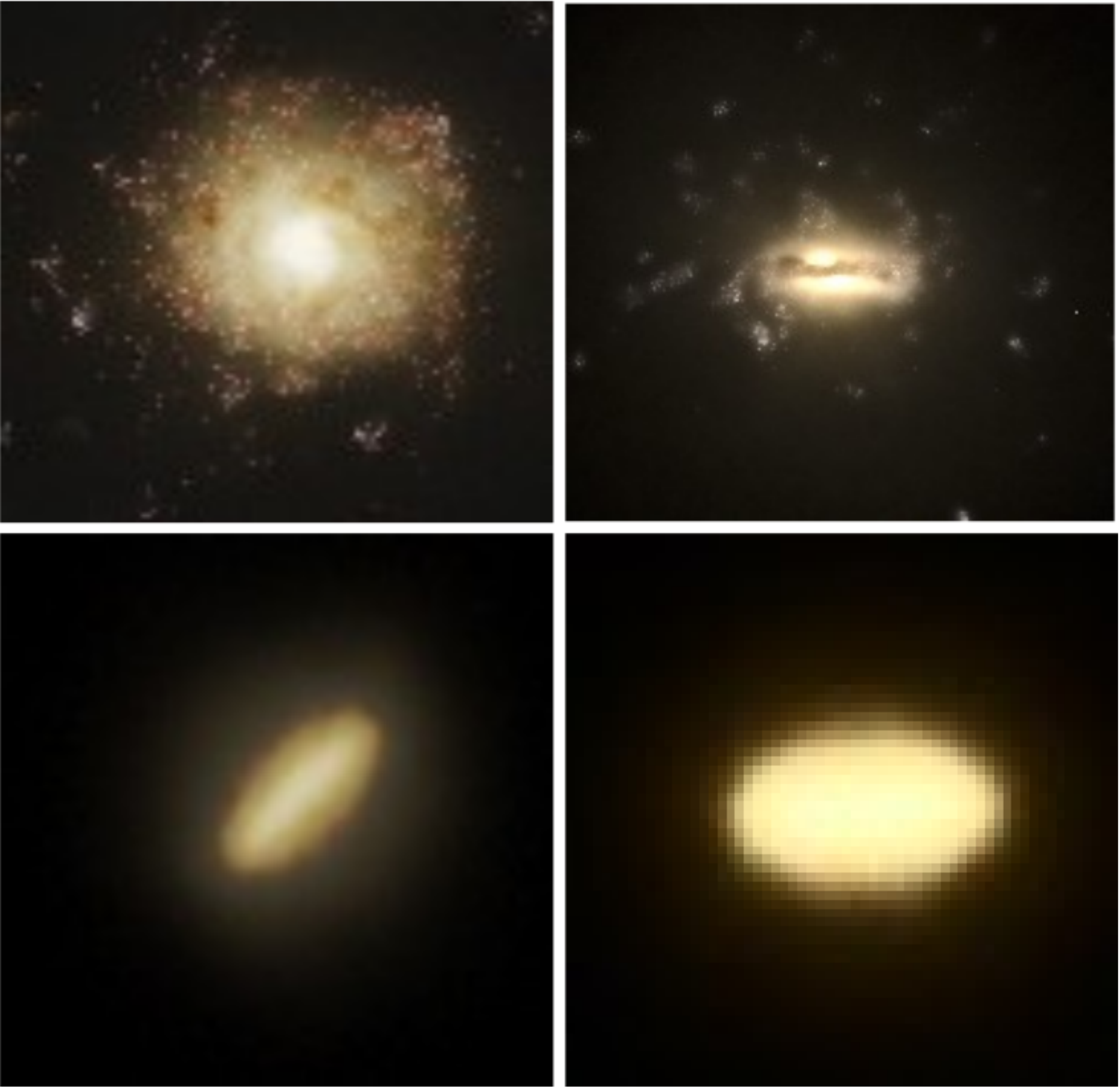}
    \caption{\textbf{Disc dominated and bulge dominated galaxies in EAGLE.} The top panel shows face-on and edge-on images of a typical spiral galaxy and the bottom panel shows the same for a typical spheroidal galaxy. 
    The stellar mass of the two galaxies are $15$ and  $5.5 \times 10^{10}$ \massUnit{}, respectively.
    The images were taken from the Eagle Public database RefL0100N1504. 
    }
    \label{fig:galaxy_image_bulge_disc}
\end{figure}

\subsubsection{Halo and galaxy spin}
\label{sec:spin}
The angular momentum or spin of a halo or galaxy is calculated by summing over the angular momentum of all the particles in it. The spin, $\mathbf{J}$, of an object with $N$ particles is given by
\begin{eqnarray}
    \mathbf{J} =  \sum\limits_{k=1}^{N} m_k \left( \mathbf{r}_{k} \times \mathbf{v_{k}}  \right)
    \;,
\end{eqnarray}
where $\mathbf{r}_k$, $\mathbf{v}_k$ and $m_k$ denote the position, velocity and mass of the $k-$th particle. The position is measured with respect to the object's centre, which is given by the most gravitationally bound particle, and the velocity is measured with respect to the centre of mass.

The DM halo spin is denoted as $\mathbf{J}_{\rm{dm}}$ and is calculated using all the DM particles within the $R_{200}$ halo radius. For galaxies we determine separately the spin of the stellar component, $\mathbf{J}_{\rm{star}}$, and of the gas disc, $\mathbf{J}_{\rm{gas}}$. The stellar spin is calculated using all the star particles within a distance of 10~kpc from the galaxy centre while the spin of the gaseous component uses all the cold (temperature below $10^5$K) gas particles within a distance of 30~kpc. In all three cases, we use only galaxies and haloes with at least 300 particles of each type.

\subsubsection{Halo and galaxy shape}
\label{sec:shape}
The shape of galaxies and haloes is usually described in terms of the ratios of the major, minor and intermediate axes. We obtain this by calculating the eigenvalues and eigenvectors of the moment of inertia tensor, 

\begin{eqnarray}
    \mathbf{I}_{ij} = \sum_{k=1}^{N} m_k r_{k,i}r_{k,j}
    \;, 
\end{eqnarray}
where $r_{k,i}$ is the position of the $k$-th particle along the $i$-th coordinate axis. The principal axes of the object are given by the eigenvectors of the $\mathbf{I}_{ij}$ tensor, $\mathbf{s}_a$, $\mathbf{s}_b$ and  $\mathbf{s}_c$, which are the directions corresponding to the major, intermediate and minor axes, respectively. 

The eigenvalues of the inertia tensor, $s_a \ge s_b \ge s_c $  are used to obtain the axis ratios $b/a$ and $ c/a$, where $a = \sqrt{s_a}$, $b = \sqrt{s_b}$ and $c = \sqrt{s_c}$. The axes ratios describe the shape of a halo. If the halo is spherical then $b/a = c/a = 1$, whereas prolate haloes have the major axis longer ($c \approx b << a $) and oblate haloes have the minor axis shorter ($c << b \approx a $) than the other two.

\subsubsection{Galaxy morphology}
\label{sec:morphology}
We classify galaxies as spheroids or discs by computing the bulge fraction $B/T$, where $B$ is the bulge mass and $T$ is the total stellar mass. The bulge mass is calculated as twice the mass of all counter-rotating stars. Specifically, if the dot product of the orbital angular momentum of a star with the total angular momentum of the galaxy is negative, then that star is considered to be counter-rotating. A galaxy which is mostly dispersion dominated will have a large fraction of counter-rotating stars, so that the value of  $B/T $ will be close to unity. If the galaxy is rotation
supported B/T is closer to zero. For our galaxy sample, over the entire galaxy mass range the median value of $B/T$, computed using star particles
within 10 kpc, is 0.76.
 
The third of the galaxy sample population with the lowest B/T ratio, ie. B/T < 0.58, are designated as disc galaxies. The third of the galaxies
with the highest $B/T$ ratio, ie. $B/T > 0.82$, are classified as spheroid galaxies. Following this classification scheme, the sample contains 2074 disc
galaxies, and an equal number of spheroid galaxies.
In \autoref{fig:galaxy_image_bulge_disc} we show representative examples of a spiral and an elliptical galaxy in the EAGLE simulation. These images were obtained from the EAGLE database and were created using the technique described in \citet{Trayford2017}.

\begin{figure*}
\vskip -0.7cm
    \subfloat{\includegraphics[width =0.95\columnwidth]
 {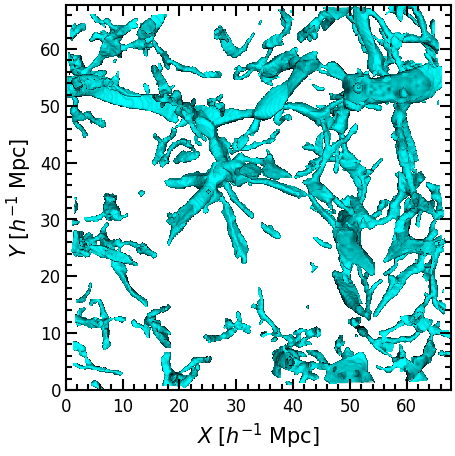}}
    \hspace{.1cm}\subfloat{\includegraphics[width =0.95\columnwidth]
 {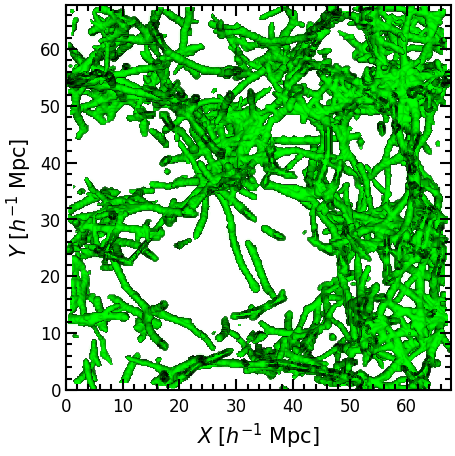}}
    \vskip -1.4cm
    \subfloat{\includegraphics[width=0.95\columnwidth]{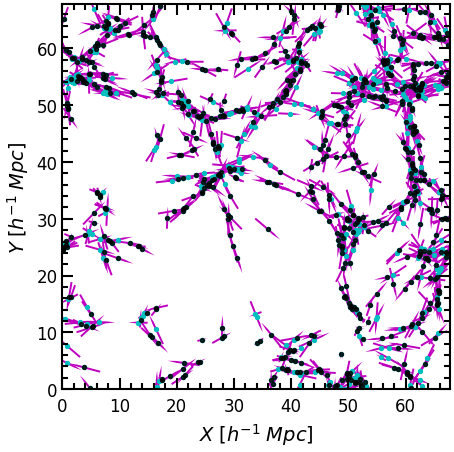}}
    \subfloat{\includegraphics[width =0.95\columnwidth]{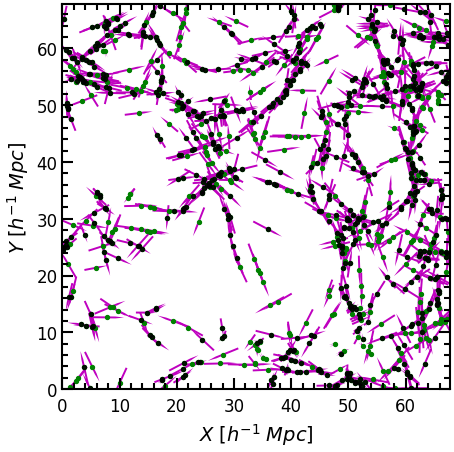}}
    \vskip -1.35cm
    \subfloat{\includegraphics[width =0.95\columnwidth]
    {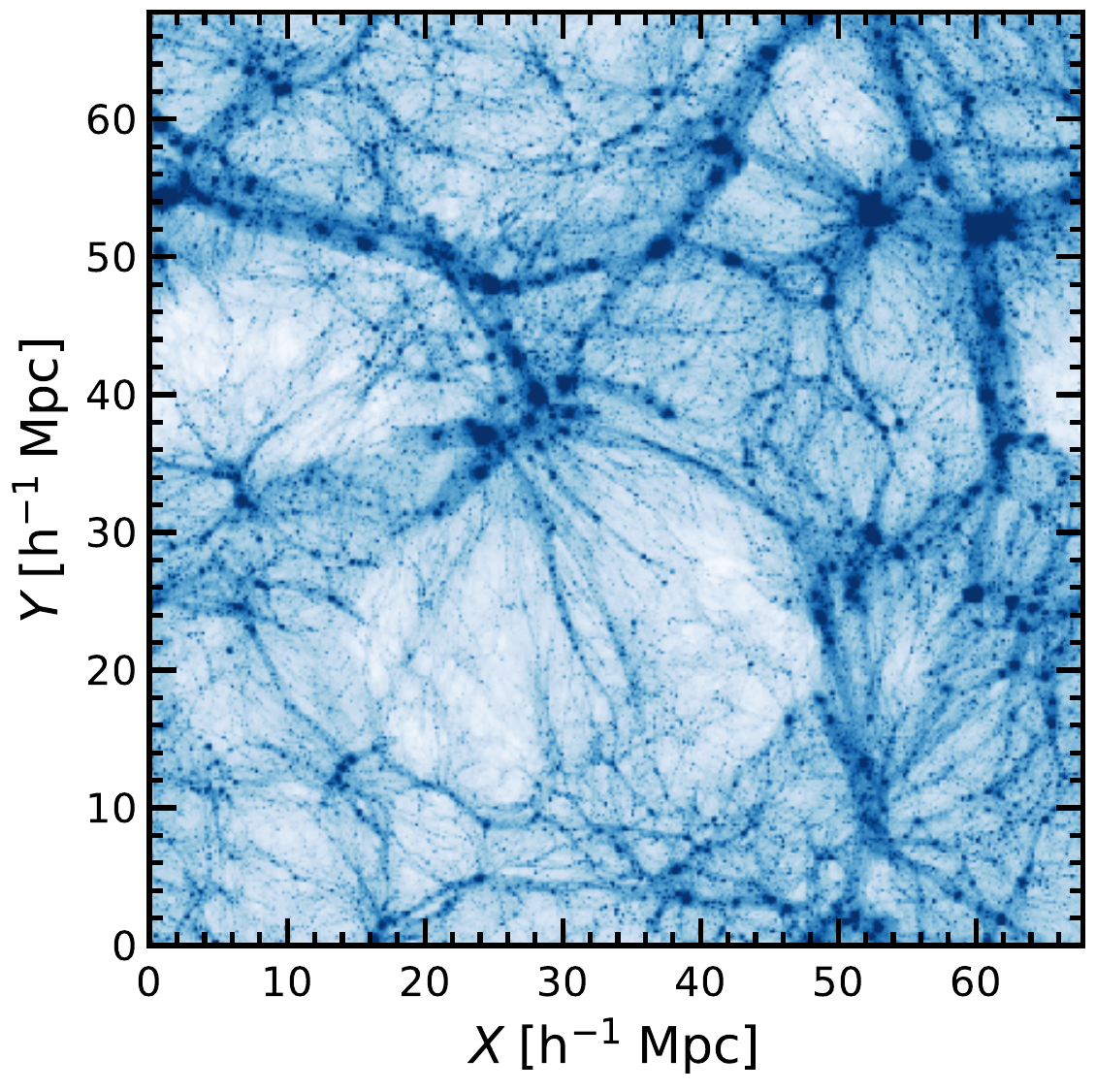}}
    \subfloat{\includegraphics[width =0.95\columnwidth]
    {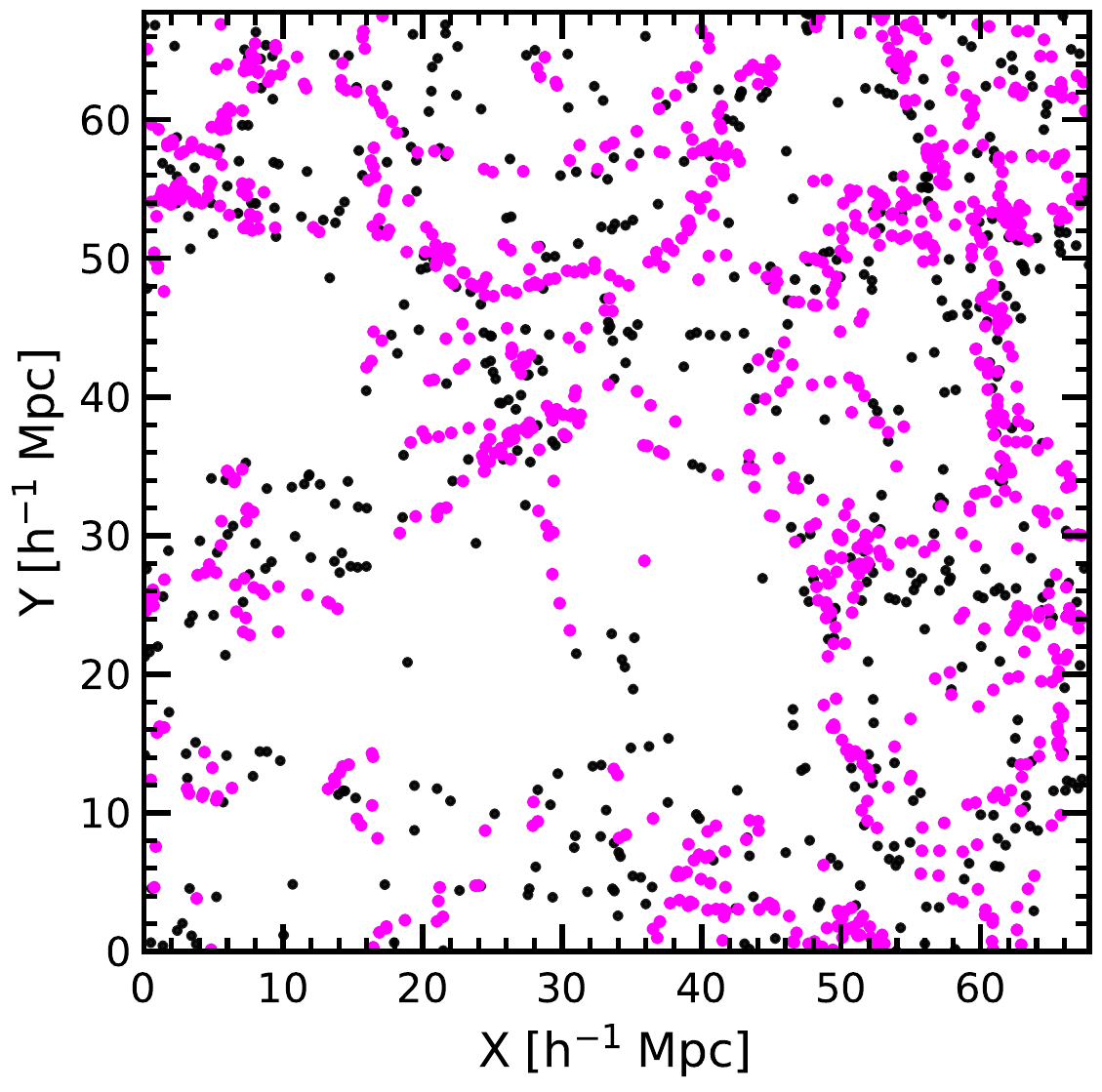}}
    \vskip -0.3cm
    \caption{\textbf{Filaments and filament galaxies in a $10$\distUnit{} slice.} \textit{Top row:} the spatial distribution of the NEXUS+ (left-hand panel) and Bisous (right-hand panel) filaments.
      \textit{Middle row:} galaxies (shown as symbols) in the NEXUS+ (left-hand panel) and Bisous (right-hand panel) filaments; the lines give the filament orientation at each galaxy position. Black dots represent galaxies common to both NEXUS+ and Bisous while blue are only in NEXUS+ and green are only in Bisous. For clarity, we show only central galaxies with stellar mass, $M_{\rm star}\ge 5 \times 10^{8}$ \massUnit{}. \textit{Bottom-left panel:} the DM density in the slice. \textit{Bottom-right panel:} all galaxies in the slice. The magenta symbols show galaxies in
      NEXUS+ filaments, the black symbols depict the rest of the galaxies.
    }
\label{fig:galaxies_in_filaments}
\end{figure*}

\vspace{-.2cm}
\section{Filament and Galaxy populations:\\ \ \ \ \ \ Nexus+ and Bisous}
\label{sec:difference_fila_population}
The Bisous algorithm uses galaxies as tracers to detect the underlying filamentary network whereas NEXUS+ uses the matter density field to identify the cosmic web. Despite this crucial difference in the tracers, the filament populations detected by both methods are almost identical with a few interesting differences that will be addressed in this section.
A visual representation of the structural features and as well as of the filament galaxy distribution is shown in \autoref{fig:galaxies_in_filaments}. 

\vspace{-.2cm}
\subsection{Structural similarities and differences}
The top two panels of \autoref{fig:galaxies_in_filaments} show the filamentary network detected by NEXUS+ and Bisous in a 10 \distUnit{} thick slice. They show that both methods identify the same overall pattern of prominent
filaments that span the weblike network pervading the simulation box and they suggest that we should expect similar halo and galaxy alignments with the two populations of filaments.

We also observe interesting differences between the NEXUS+ and Bisous filaments related to the thickness of individual structures. NEXUS+ filaments have a range of thicknesses, while all Bisous objects have roughly the same thickness. This contrast is related to differences in the formalism underlying the methods. NEXUS+ is an explicit multiscale method and belongs
to the Multiscale Morphology Filter/Nexus family of cosmic web classification tools \citep{aragon2007MMF,cautun2013}. In the
implementation for the present study we used smoothing scales ranging  from 0.5 to 4 \distUnit{}. The panels in
\autoref{fig:galaxies_in_filaments} reflect this: the NEXUS+ filaments vary in thickness, ranging from very thin to very thick. 

In contrast, the Bisous formalism identifies filaments using a fixed transverse filament scale of 1$\rm{Mpc}$ (0.68\distUnit{}), which translates
into cylindrically shaped filaments with a radius of 1$\rm{Mpc}$. As a result, we see a few heavy and thick NEXUS+ filaments
that correspond to a configuration of parallel cylindrical Bisous filaments. Note that the orientation of the thick NEXUS+ and the
Bisous filaments will be largely similar.  Also, we see a substantial difference in the identification and classification of
small scale tenuous filaments, in particular in moderate and lower density regions. Several of the smaller Bisous filaments 
located in these regions are embedded in regions that NEXUS+ assigns to walls and voids. We see this illustrated in the
central region and the bottom righthand corner of the panel showing the Bisous filamentary network in
\autoref{fig:galaxies_in_filaments}, where many Bisous filaments crisscross to form sheet-like structures. This is a consequence of
the focus of the Bisous formalism on fixed radius and elongated cylindrical features. 

The structural differences between the NEXUS+ and Bisous filaments are also reflected in a quantitive comparison
of the mass and volume filling fractions of their DM, gas, and galaxy content. For this 
complete inventory we refer to \autoref{table:mass_volume_table}. The fractions were calculated by splitting the EAGLE box into a $256^3$ grid (grid spacing of $0.26$ \distUnit{}) and counting the number of grid cells associated to each cosmic web component. We
find a reasonable agreement between the two filament populations, with some modest differences. The NEXUS+ filaments contain a
slightly higher DM, gas and stellar mass fractions than Bisous filaments. This result, which is consistent with the one reported in
\citet{Libeskind2018}, is probably a reflection of the fact that prominent NEXUS+ filaments are substantially thicker than their Bisous
counterparts and thus contain more of the cosmic mass budget.

\begin{figure}
 	\includegraphics[width=\columnwidth]{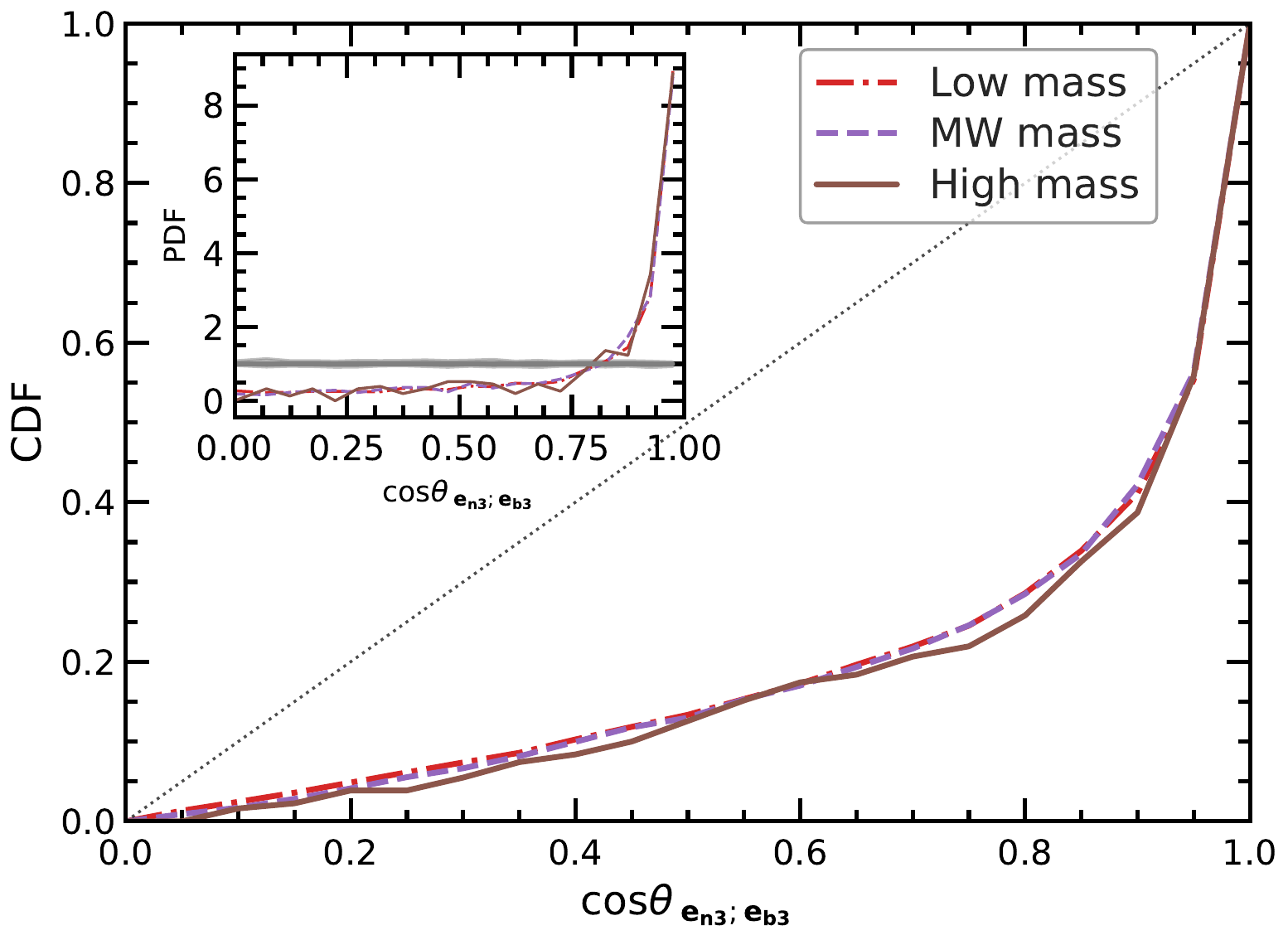}
 	\vskip -.2cm
	\caption{\textbf{Alignment between NEXUS+ and Bisous filaments.} The graph shows the CDF and PDF (inset) of the alignment angle between the orientation of NEXUS+ ($\mathbf{e}_{n3}$) and Bisous ($\mathbf{e}_{b3}$) filaments. The alignment is measured at the position of common filament galaxies. The various coloured lines correspond to galaxies of different stellar masses: low mass, $\leq 1 \times 10^{10}$ \massUnit{}, intermediate mass, $(1 - 5) \times 10^{10}$ \massUnit{}, and high mass, $\geq 5 \times 10^{10} $ \massUnit{}. Irrespective of the galaxy mass range, NEXUS+ and Bisous filaments are well aligned with each other.
	}
    \label{fig:cdf_fila_nex_bis}
\end{figure}

Given the focus of our study on the alignment of haloes and galaxies with their host filaments, it is crucial to compare the orientations of the Bisous and of the NEXUS+ filaments. To this end, in \autoref{fig:cdf_fila_nex_bis}
we plot the cumulative distribution of the cosine of the angle between the  third eigenvector of Bisous ($\mathbf{e_{b3}}$) and of NEXUS+ ($\mathbf{e_{n3}}$)
filaments. For an objective comparison, we  assess the mutual orientation of the  filaments  at the locations of common galaxies that are assigned to filaments by
both the Bisous and the NEXUS+ methods. Overall, we find a high degree of alignment between the two filament
populations with a median alignment angle of ${\sim}21^{\circ}$. There is
also no noticeable dependence of the alignment angle on galaxy mass, with the alignment distribution for high and low mass galaxies being practically indistinguishable. 

\begin{table}
    \centering
    \caption{\textbf{Number of EAGLE galaxies found in filaments.} The table gives the galaxy counts with stellar mass, $ M_{\rm star} \geq 5 \times 10^{8}$~\massUnit, residing in NEXUS+ and Bisous filaments. In total, the EAGLE simulation contains 9563 galaxies more massive than the above stellar mass cut. 
    The third column shows the fraction of galaxies found in the two filament populations. The fourth columns gives the number of galaxies in common to both NEXUS+ and Bisous filaments, while the last columns gives the number of exclusive filament galaxies, that is those assigned to filaments by one method but not by the other one.} 
    \label{table:number_galaxies}
	\begin{tabular}{ @{}l ccccc@{}} 
		\hline
        \\[-0.5cm]
        \hline
		Filaments & Total & Fraction & Overlap & Exclusive  \\
         & & [$ \% $ ]& & \\
        \hline
		NEXUS+ & $6394$ & 66.9 & $ \multirow{ 2}{*}{$4277$} $ & $2117$  \\
		Bisous & $5988$ & 62.6 &$ $ & $1711$ \\
		\hline 
    \end{tabular}
\end{table}

\bigskip
In summary, the NEXUS+ and the Bisous web finders both detect the major prominent filamentary arteries of the cosmic web, however, there are substantial differences between the methods in the population of small-scale filaments. The multiscale nature of
NEXUS+ allows it to detect filaments of different widths, while Bisous concentrates on filaments of a particular specified scale. Of considerable importance for this study is that the common Bisous and NEXUS+ filaments are well aligned with respect to each other. 

\subsection{Galaxy distribution in filaments: NEXUS+ vs. Bisous}
\label{subsec:differences_galaxyPopulation}
In addition to the structural characteristics discussed above, an important aspect of filament properties concern their galaxy population. 
Here we compare the galaxy populations in NEXUS+ and Bisous filaments, with \autoref{table:number_galaxies} giving an overview of
the number and fraction of EAGLE galaxies located in filaments. We limit the analysis to 
galaxies with a stellar mass in excess of $ M_{\rm star} \geq 5 \times 10^{8}$~\massUnit{}, which are the ones resolved with enough particles to have robust spin and shape measurements.

NEXUS+ filaments contain $67\%$ of the total number of (central) galaxies, while Bisous identifies a slightly lower fraction of filament galaxies, $63\%$. As we discuss in \autoref{sec:mass_functions}, an important difference between the two web finders is that Bisous assigns a considerably lower fraction of massive galaxies (that is those with $M_{\rm star} \geq  10^{11}$~\massUnit{}) to filaments than NEXUS+. These massive galaxies are usually located in the nodes of the cosmic web and in their immediate neighbourhoods, which are regions that Bisous does not classify as filaments (see the discussion in \autoref{sec:mass_functions}).

Of the entire NEXUS+ population of filament galaxies, $~67\%$ of them are residing in Bisous filaments. Meanwhile, some $71\%$ of Bisous filament galaxies are also found in NEXUS+ filaments. The rest of the Bisous galaxies are in regions classified as walls ($27.5\%$), voids ($0.85\%$) and clusters ($0.16\%$) by NEXUS+. In short, the majority of Bisous filament galaxies are also located in NEXUS+ filaments, although a considerable fraction appears to be located
in regions identified as walls by NEXUS+. 

\begin{figure}
    \centering
    \includegraphics[width =\columnwidth]{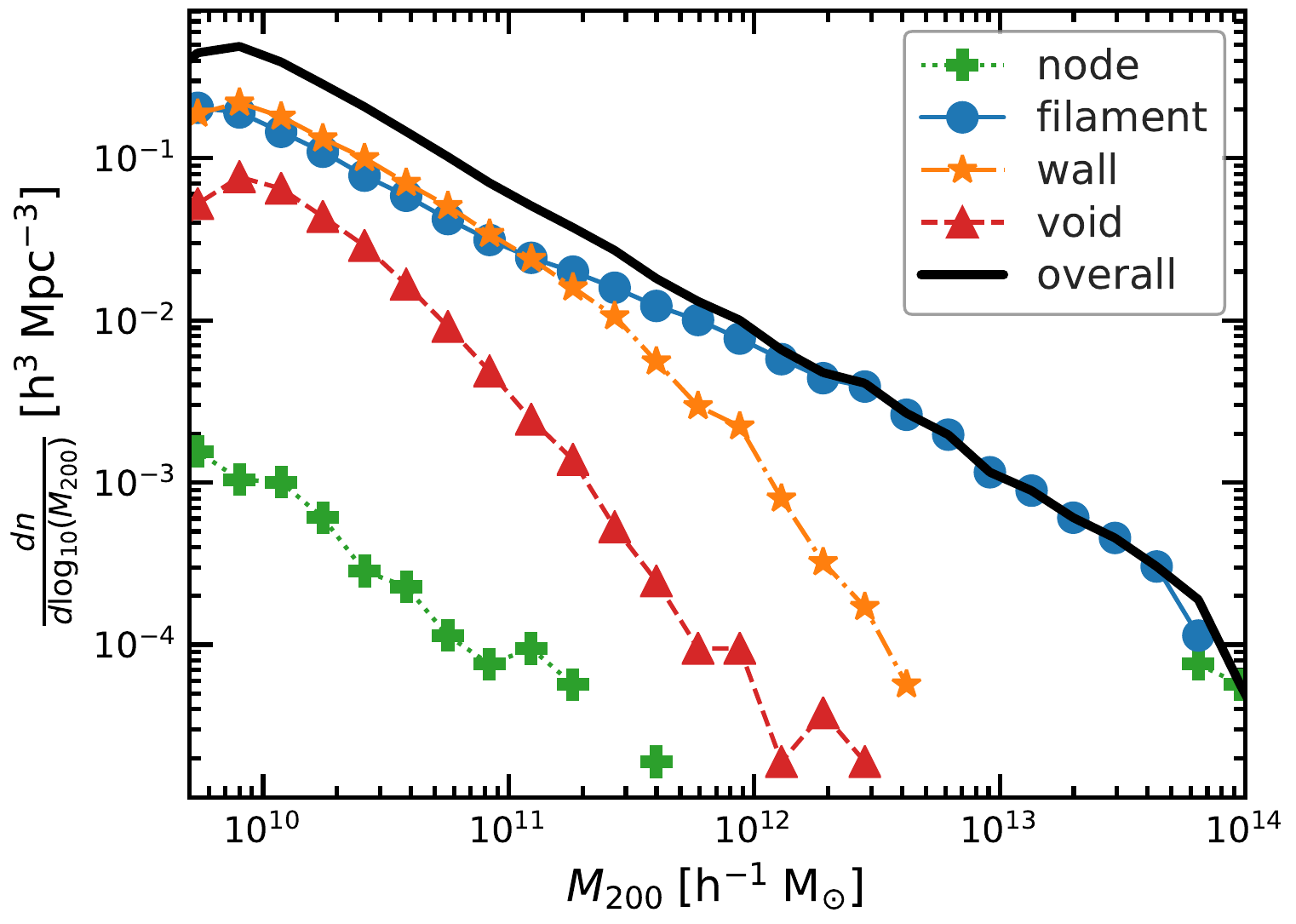}
    \vskip -0.1cm
    \includegraphics[width =\columnwidth]
    {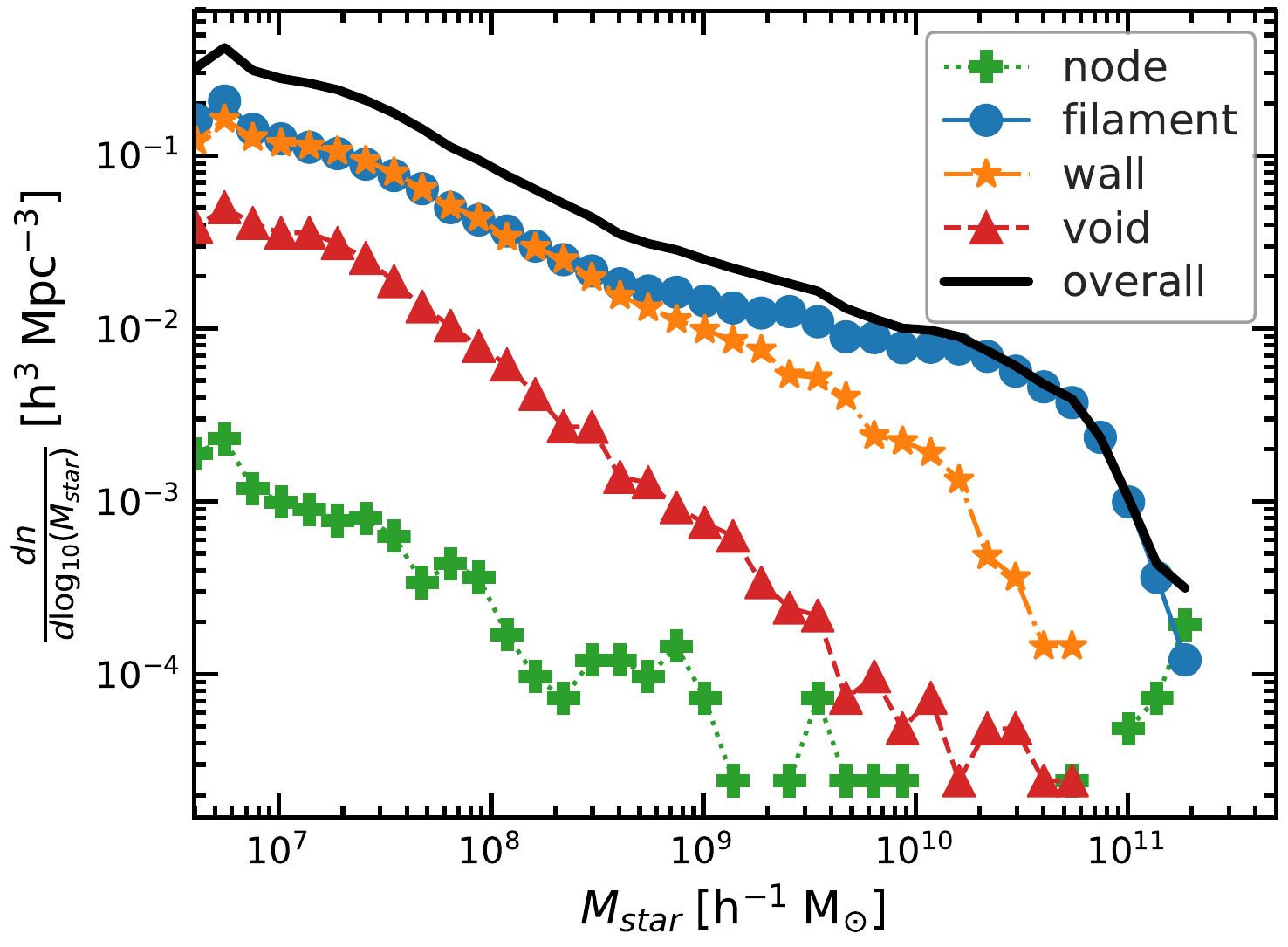}
    \vskip -0.1cm
    \includegraphics[width =\columnwidth]
    {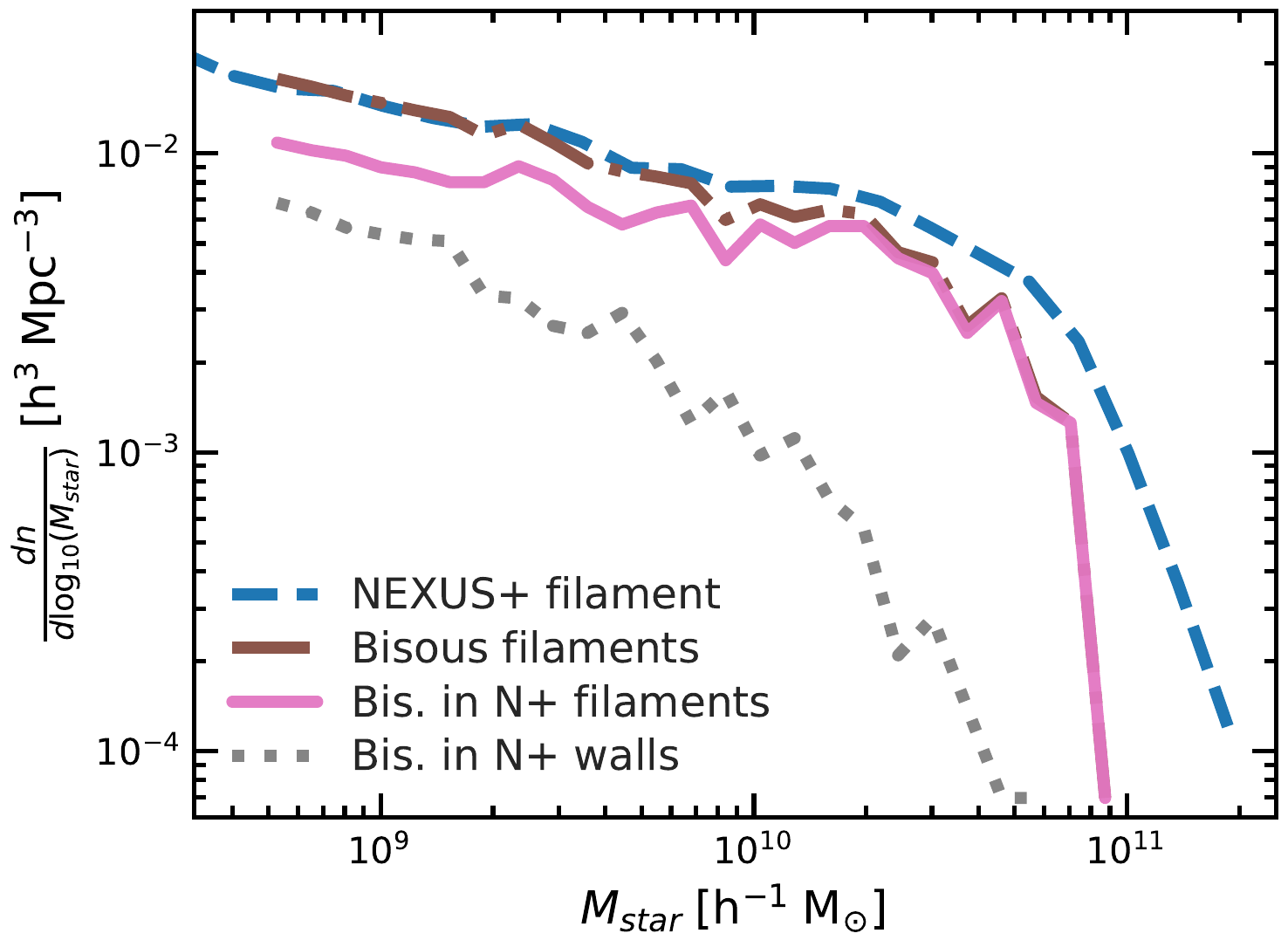}
    \vskip -.3cm
    \caption{\textbf{Halo and galaxy stellar mass functions segregated by web environment.} The top panel shows the halo mass function in NEXUS+ environments. The centre panel shows the galaxy mass function in NEXUS+ environments. The bottom panel compares the galaxy mass function in NEXUS+ (dashed line) and Bisous (dashed-dotted line) filaments. The Bisous filament galaxies are mostly found in NEXUS+ filaments (solid line) and, a small fraction of them, in NEXUS+ walls (dotted line).
    }
    \vskip -.2cm
    \label{fig:halo_mass_function}
\end{figure}

A visual appreciation of the spatial distribution of galaxies in the cosmic web can be obtained from the two central row panels in \autoref{fig:galaxies_in_filaments}. They show the filament galaxy population of the two web finders, with common galaxies associated to both NEXUS+ and Bisous filaments shown as black symbols. The  most outstanding difference concerns the galaxies populating the thin filaments, which are typically low mass galaxies. While the number of
low mass filament galaxies is comparable, the low mass galaxies that are not shared by Bisous and NEXUS+ often concern the ones that have
been classified as wall galaxies by NEXUS+.

\subsection{Halo \& galaxy mass functions}
\label{sec:mass_functions}
\autoref{fig:halo_mass_function} plots the halo and galaxy stellar mass functions of the EAGLE simulation as a function of
the cosmic web environment, i.e. the mass functions of dark halos and galaxies in the nodes, filaments, walls and voids of the cosmic web. To this
end, we plot the number density of haloes and galaxies per logarithmic mass bin. The
top panel shows the halo mass function split into web environments as determined by NEXUS+. The corresponding galaxy stellar mass function is given in the central panel.
The bottom panel compares the galaxy stellar mass function identified in Bisous filaments and with that assigned to the NEXUS+ filaments. We mostly limit our analysis to galaxies with $M_\mathrm{star} \ge 5  \times 10^{8} \; h^{-1} M_{\odot}$, which represents the population of objects whose spin--filament is the topic of this paper. However, for completeness, in the case of NEXUS+ environments we show the galaxy mass function down to much fainter central galaxies with $M_\mathrm{star} \ge 5  \times 10^{6} \; h^{-1} M_{\odot}$ (corresponding to roughly 3 or more star particles). 

The first two panels of \autoref{fig:halo_mass_function} show that a majority of haloes with a mass,
$M_{200} \ge 5 \times 10^{11}$ \massUnit{}, and galaxies with a stellar mass, $M_\mathrm{star} \ge 5  \times 10^{9}$ \massUnit{}, are located in the filaments of the cosmic web
\citep{cautun2014,Libeskind2018}. In this mass range, walls, and even more so voids,
represent considerably more desolate environments. These results are in good agreement with observational studies, such as \citet{Tempel2011} and
\citet{Eardley2015}, which show that the galaxy luminosity function varies between different environments. More specifically, \citeauthor{Eardley2015} find that
the number density of galaxies as well as the knee of the Schechter function used to fit the galaxy luminosity function \citep{Schechter1976} are
the highest for nodes and decreases going from filament, to wall and to void environments.

We find a similar trend for galaxies in filaments, walls and voids but not for nodes. Due to its small box size, which is only $100$~Mpc on a side,
the EAGLE simulations is not able to produce a representative population of massive cluster sized
haloes. For a structure to be identified as a node by the NEXUS+ algorithm, we use a mass threshold of $5\times 10^{13}$\massUnit{}. Owing to the truncated power spectrum due to the small box size, such massive structures are not formed in the EAGLE simulation. It translates into a
substantial suppression of the halo and galaxy mass function in the cosmic web nodes, in line with the finding of eg. \citet{Bagla2005}.
They already showed that the high mass end of the halo mass function is significantly reduced in a $\Lambda$CDM simulation
volume with a side length less than $100$~\distUnit{}.

The differences between the NEXUS+ and the Bisous filaments are also reflected in the corresponding filament galaxy mass functions. This are illustrated in the bottom panel of \autoref{fig:halo_mass_function} which compares the galaxy mass function of Bisous and NEXUS+ filaments. At the low mass of the mass function, both web finders assign a similar number of galaxies to filaments. At the high mass end, we see a marked difference. While NEXUS+ assigns a range of massive galaxies to filaments, the Bisous formalism yields a sharp cutoff at $M_{star} \sim10^{11}$ \massUnit{}. Such differences are not uncommon, as we may infer from the detailed comparison in \citet{Libeskind2018}, so actually the agreement between the galaxy mass function in NEXUS+ and Bisous filaments is rather good. One factor that might lead to enhanced differences between the two web finders is the very low number of massive clusters in the EAGLE simulation. On one hand, this results in NEXUS+ assigning more massive galaxies to filaments. On the other hand, the Bisous selection criteria related to the strength and orientation of valid filaments refrain the Bisous filaments from extending close to the high density nodes. Filaments detected next to cosmic web nodes have a lower orientation strength and therefore galaxies surrounding the nodes, which are predominantly more massive, might not be part of the Bisous filamentary network.

Regardless of the differences discussed above, we find that the majority of Bisous filament galaxies are also identified as filament galaxies by NEXUS+. This can be inferred from the solid line in the bottom panel of \autoref{fig:halo_mass_function}, which shows the galaxies common to both Bisous and NEXUS+ filaments. A small fraction of Bisous galaxies turn out to be associated with NEXUS+ walls, while a minute number is found in either nodes or voids.

\vskip -.2cm
\section{Alignment analysis and results}
The present section presents our results on the alignment of the spin and shape of haloes and galaxies with respect
to the orientation of the large-scale filaments in which they reside. We assess the
cosmic web alignment on the basis of four different aspects:
\begin{itemize}
\item[$\bullet$] the alignment of the spin of haloes and the spin of galaxies with respect to the
filament to which they are bound (sect.~\ref{sec:halo_gal_spin_alignment}).
\item[$\bullet$] the differences between the alignment of late-type disc galaxies to the filament in which they reside
  and that of early-type galaxies (sect.~\ref{sec:alignment_morphology}).
\item[$\bullet$] the observationally more accessible alignment of galaxy and halo shape, in terms of
  their minor axis, with respect to the embedding filament (sect.~\ref{sec:shape_alignment}).
\item[$\bullet$] the alignment between the spins of galaxies and their filaments, as well as that between the minor axis of galaxies and their haloes (sect.~\ref{sec:halo_galaxy}). 
\end{itemize}

\begin{figure}
    \includegraphics[width=\columnwidth]{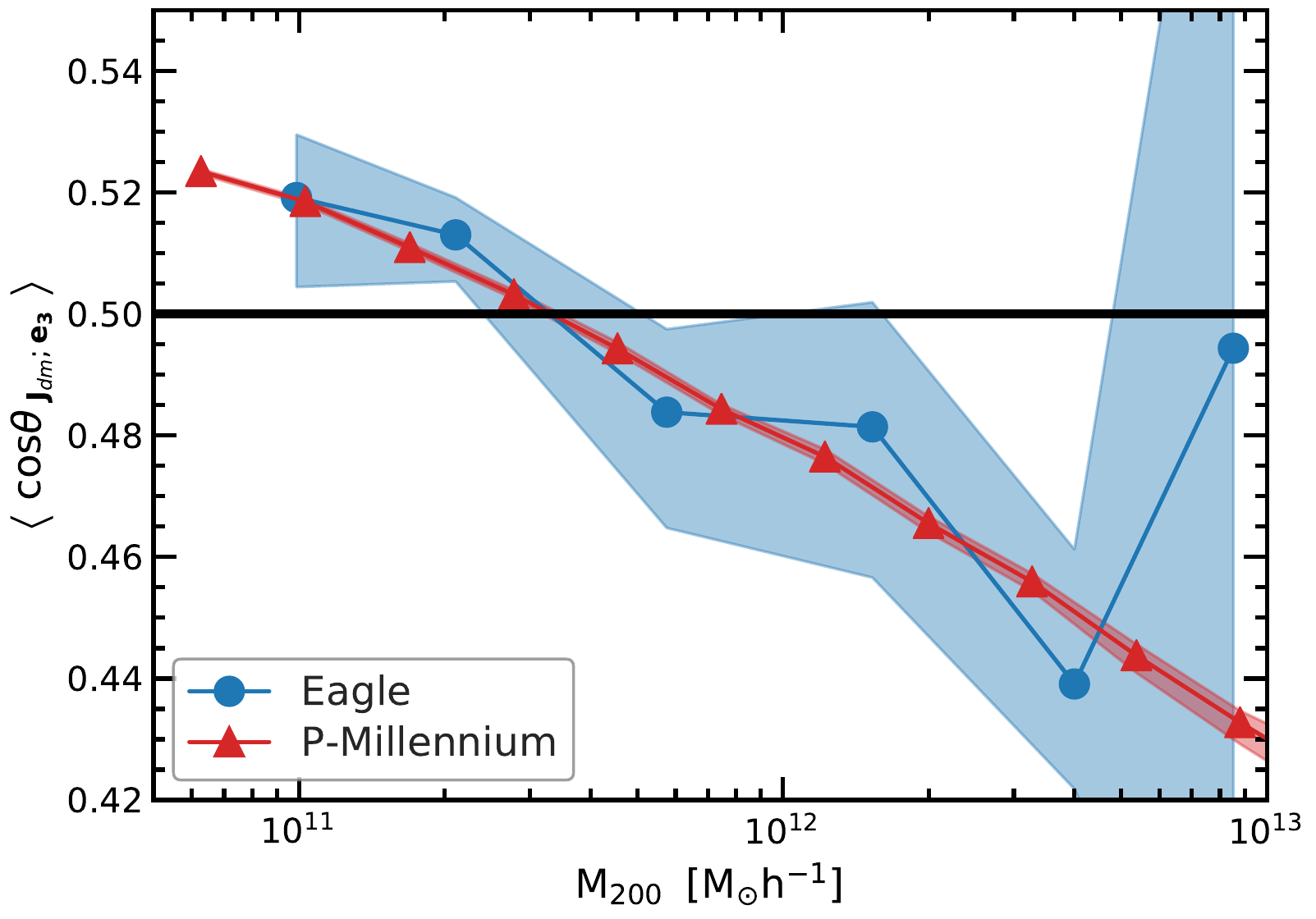}
    \vskip -.2cm
    \caption{\textbf{Halo spin -- filament alignment.} The dependence on halo mass of the median alignment angle
    between the spin of DM haloes and the orientation of NEXUS+ filaments. The plot compares the alignment in the EAGLE hydrodynamical simulation with the one in the P-Millennium DM-only simulation.
    }
    \label{fig:median_eagle_pmillennium}
\end{figure}

\begin{figure*}
    \includegraphics[width=\textwidth]{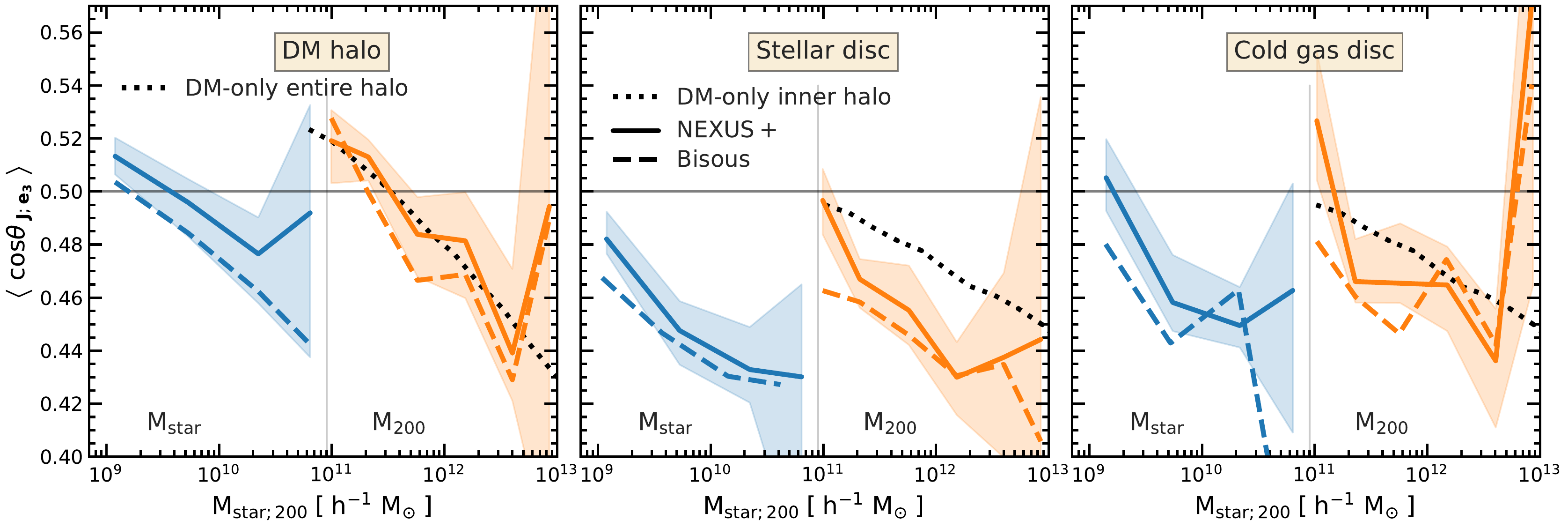}
    \vskip -.1cm
    \caption{\textbf{Spin--filament alignment for haloes and galaxies}. The alignment is plotted as a function of halo mass (orange) and stellar mass (blue) for central galaxies and their host haloes. The left panel is for host haloes, the central panel is for the stellar disc and the right panel for the cold gas disc. In all the panels the solid line shows the alignment with NEXUS+ filaments and the dashed line the alignment with Bisous filaments. The black dotted line shows the alignment of the entire halo in the left panel and the inner halo in the centre and right panels measured in the P-Millennium DM-only simulation. The shaded region represents the $2\sigma$ uncertainty and is plotted only for NEXUS+ filaments for clarity. The error range for Bisous is very similar.
    }
    \label{fig:median_angular_momentum}
\end{figure*}
\begin{figure}
    \includegraphics[width=\columnwidth]{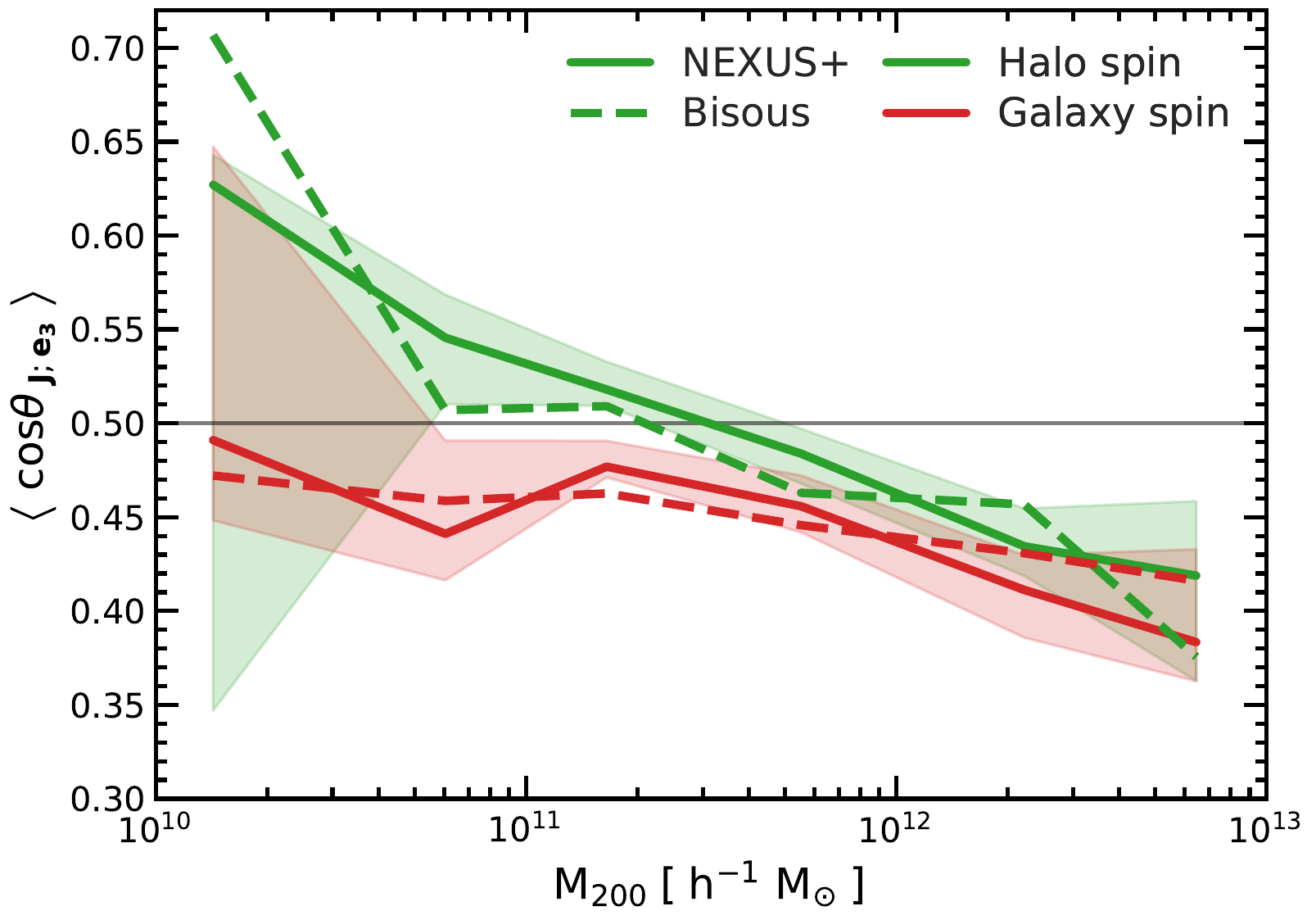}
    \vskip -.2cm
    \caption{\textbf{Median spin alignment of common galaxies.} The plot shows the spin--filament alignment only for the galaxies and the haloes common to both NEXUS+ and Bisous filaments. The consistent difference seen in \autoref{fig:median_angular_momentum} between spin alignments with NEXUS+ and Bisous filaments does not exist when we choose galaxies that are common to both type of filaments.
}
\label{fig:common_bisous_nexus}
\end{figure}

\begin{figure*}
    \includegraphics[width=\textwidth]{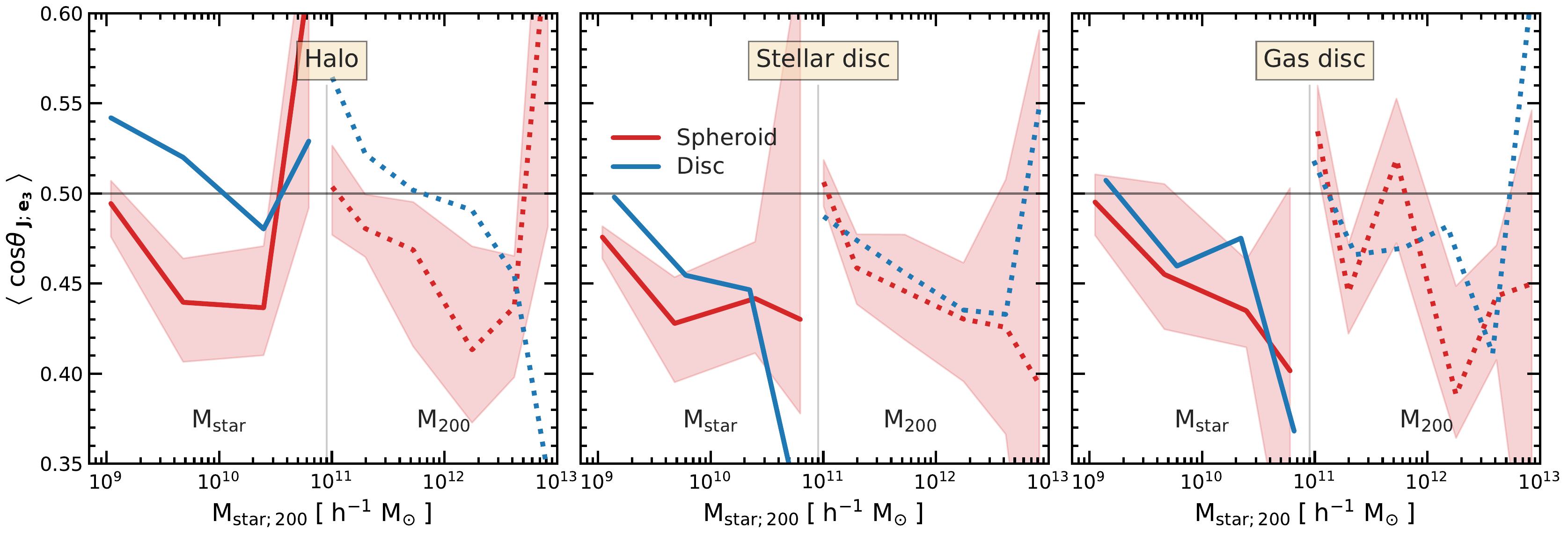}
    \vskip -.1cm
    \caption{\textbf{Spin--filament alignment for discs and spheroids.} The plot shows the alignment as a function of both stellar mass (solid line) and halo mass (dotted line) for disc ($B/T < 0.58$, shown in blue) and spheroid ($B/T > 0.82$, shown in red) galaxies. The disc and spheroid samples were selected to contain a third of the galaxy population with respectively the lowest and highest bulge to total ratio. We only show the alignment with NEXUS+ filaments; however this is very similar for Bisous filaments. The shaded region shows the uncertainty, as in \autoref{fig:median_angular_momentum}. 
    }
    \label{fig:median_morphology}
\end{figure*}

\subsection{Spin and shape alignment analysis}
\label{sec:alignment_analysis} 
In order to quantify the alignment between galaxies and haloes on the one hand and the filaments in which they reside, on the other hand, we define the misalignment angle, $\theta$, as the angle between two vectors, one of which corresponds to the property of a halo or galaxy
($\mathbf{h}$), and the other corresponds to filament orientation ($\mathbf{e_3}$), which is the slowest collapse direction. The alignment parameter, or simply the alignment angle, is given by 
\begin{eqnarray}
    \mu_{hf}\, \equiv \,
    \cos \theta_{\mathbf{h}, \mathbf{e}_3} = \left| \frac{\mathbf{h} \cdot \mathbf{e}_3}{|\mathbf{h}| |\mathbf{e}_3 |} \right| \,,
    \label{cos}  
\end{eqnarray}
where we take the absolute value of the scalar product since filaments have an orientation, but not a direction. That is, both $\mathbf{e}_3$ or $-\mathbf{e}_3$ point along the filament axis. A vector quantity that is parallel to the filament axis (either to $\mathbf{e}_3$ or $-\mathbf{e}_3$), corresponds to $\mu_{hf} = 1$. When the galaxy spin, or shape, is directed perpendicular to filaments, it yields an alignment parameter, $\mu_{hf} = 0$. 

In general, the halo/galaxy vector properties have a distribution of alignment angles with respect to the filament axis. We can quantify this using the probability distribution function (PDF) of the alignment angle. Furthermore, the PDF can vary according to halo/galaxy mass (see e.g. Figure 10 in \citealt{GVeena2018}), and thus different mass sample can have different PDF distributions. To quantify this mass dependence, we calculate the median alignment parameter, $\left\langle\mu_{hf}\right\rangle$, as a function of halo and galaxy mass. In the absence of any alignment, that is in the case of an isotropic distribution of alignment angles, the PDF of $\mu_{hf}$ is an uniform distribution between 0 and 1, and has a median value, $\left\langle\mu_{hf}\right\rangle=0.5$. 
To calculate the alignment angle uncertainties, we generate many bootstrap realizations for each mass bin. From these, we estimate the $1 \sigma $ and $2 \sigma$ uncertainty intervals for both the PDF and the median alignment angle (see e.g. \autoref{fig:median_eagle_pmillennium}). 

We characterise a population to be preferentially parallel if the median alignment is $\left\langle\mu_{hf}\right\rangle > 0.5$. Conversely,  $\left\langle\mu_{hf}\right\rangle < 0.5$ corresponds to a  preferentially perpendicularly aligned population. The alignment parameter $\mu_{hf} = 0.5$ marks the transition between preferentially parallel and perpendicular and it corresponds to an angle, $\theta_{h,f} =  60^{\circ}$.

\subsection{Halo and galaxy spin--filament alignment}
\label{sec:halo_gal_spin_alignment} 
We first study the alignments between the DM halo spin and its host filament. This has been extensively studied in DM-only simulations (see discussion in the introduction section) and we want to assess if the inclusion of baryonic physics affects this alignment. \autoref{fig:median_eagle_pmillennium} compares the median halo spin--filament alignment angle as a function of halo mass in two simulations: EAGLE, which includes galaxy formation processes, and the P-Millennium DM-only simulation. The latter is a very high resolution, $5040^3$ DM particles each of mass $1.061 \times 10^{8}~{h}^{-1}\rm{M}_{\odot}$, and large volume, a $800$ Mpc periodic box, simulation \citep{McCullagh2017,Baugh2018} of structure formation in a $\Lambda$CDM universe with the same cosmological parameters as the EAGLE simulation. \citet{GVeena2018} have used the very large sample of P-Millennium haloes to characterize the halo spin--filament alignment over a wide range of halo masses; their result is the one shown in \autoref{fig:median_eagle_pmillennium}. Note that for both the EAGLE and the P-Millennium simulations we use the same filament finding algorithm, NEXUS+, in order to eliminate discrepancies arising from the use of different web finders.  

Despite differences in the initial conditions, box size and the nature of the simulations, the halo spin--filament alignments in P-Millennium and EAGLE are statistically identical.
In both simulations, the spin of less massive haloes shows a preferential parallel alignment with the filaments whereas the spin of massive haloes shows a preferential perpendicular alignment. The mass where this transition happens is known as the spin-flip mass and is identical in both simulations. Thus, the inclusion of baryons does not alter (at least given the statistics of the EAGLE sample) the mean alignment between halo spins and their host filaments.

Next, we study the galaxy spin--filament alignment and compare it to the mass dependent alignment trend seen for haloes.  
In \autoref{fig:median_angular_momentum} we show the median alignment of haloes and stellar and gaseous components of galaxies. The left-hand panel shows the halo spin--filament alignment: haloes show a clear spin transition from preferentially parallel to perpendicular with respect to both halo mass and the stellar mass of their central galaxies. The  stellar (centre panel) and gas (right-hand panel) components also show a strong mass dependent alignment, with high mass galaxies showing a stronger perpendicular alignment than low mass galaxies. The spin alignment of the stellar and gas distributions are very similar, although the gas spin is slightly less perpendicular to filaments than the stellar component. However, neither the stars nor the gas components exhibit a spin transition as in the case of their host haloes.

This discrepancy could be due to differences in angular momentum acquisition between haloes and galaxies. For example, \citet{GVeena2018} have pointed out that the halo spin--filament alignment varies between the inner and full halo. At fixed halo mass, the spin of the inner halo is more orthogonal to the filament spine than that of the full halo. 
Galaxies are likely to be better aligned with the inner regions of their host haloes \citep[e.g.][]{Bailin2005,Velliscig2015,Shao2016}, which motivates us to compare with the spin--filament alignment of the inner regions of haloes in DM-only simulations. For this, we use the \citeauthor{GVeena2018} results for the spin--filament alignment of the inner 10\% of the DM halo mass, which are shown as a dotted line in the centre panel of \autoref{fig:median_angular_momentum}. It shows a closer match to the galaxy spin--filament alignment, although there are still discrepancies: the galaxy spin in EAGLE is systematically more orthogonal to the host filament axis than the spin of the inner halo in DM-only simulations. Because of feedback and dissipation processes the angular momentum build-up in galaxies can be different from that of the inner halo regions and thus stellar spin can deviate somewhat from that of the inner halo. 

\bigskip
We also investigate whether the alignment trend is sensitive to the tracers and techniques used for filament detection. \autoref{fig:median_angular_momentum} also compares the median spin--filament alignment for two different filament populations identified using the NEXUS+ (solid lines) and Bisous (dashed lines) methods. Despite the various differences listed in \autoref{sec:difference_fila_population}, the alignment trend is robust and consistent irrespective of the filament type. However, we find that galaxies and haloes in Bisous show a consistently stronger orthogonal signal than NEXUS+ filaments. 

Previous studies have shown that the alignment varies strongly with the properties of filaments, with haloes found in thinner filaments having their spin more perpendicular on the filament axis than equal mass haloes in thicker filaments \citep{aragon2014,GVeena2018}. Could the same phenomenon explain the differences in halo and spin alignment between NEXUS+ and Bisous filaments? We investigate this in \autoref{fig:common_bisous_nexus}, where we plot the median alignment using only the sample of haloes and galaxies common to both Bisous and NEXUS+ filaments. The figure shows that the alignment of the common sample with the two filament types is statistically indistinguishable and that there is no systematic discrepancy. It is interesting to note that most of the common galaxies are the ones located in prominent and dynamically active filaments (see black symbols in the central row of panels in \autoref{fig:galaxies_in_filaments}).  Thus, the differences seen in \autoref{fig:median_angular_momentum} are mostly due to non-overlapping galaxies found either in the peripheral regions of filaments or in tenuous filaments. Motivated by this find, we checked if the galaxy spin--filament alignment varies with filament properties, such as thickness, and, while we found a hint of such a trend, the EAGLE simulation does not have a large enough sample of galaxies to robustly claim such a dependence.

\begin{figure*}
    \includegraphics[width=\textwidth]{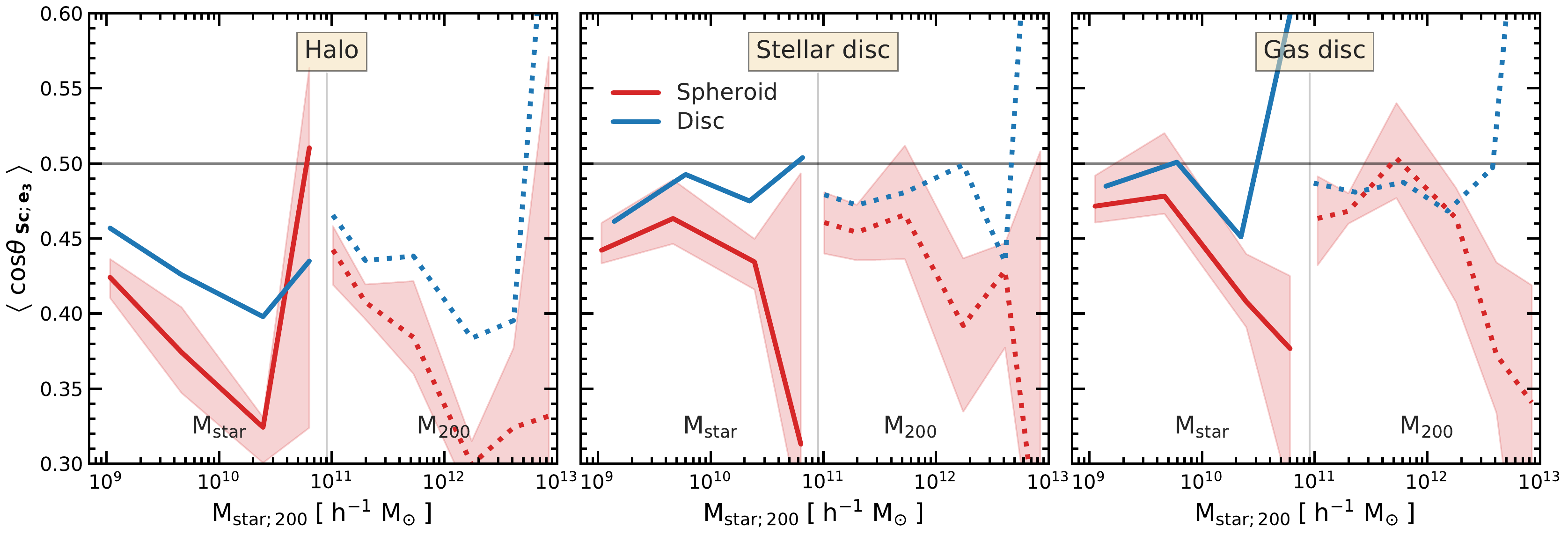}
    \vskip -.2cm
    \caption{\textbf{Galaxy shape--filament alignment.} Same as \autoref{fig:median_morphology} but for the median alignment angle between the filament orientation and the minor axis of the halo (left panel), stellar (centre panel) and gas (right panel) components. The sample is divided into systems for which the central galaxy is either disc- (blue) or bulge-dominated (red). }
    \label{fig:median_shape_filament}
\end{figure*}

\subsection{Spin alignment and galaxy morphology}
\label{sec:alignment_morphology} 
It has been shown observationally that spirals and spheroids show different alignments with their host filaments. Spirals are typically less perpendicular to the filament axis than spheroid galaxies \citep[e.g. see][]{tempel2013}. This motivated us to study how the galaxy spin--filament alignment in EAGLE varies with galaxy morphology. For this, we divide the population into disc- and bulge-dominated galaxies based on the bulge to total ratio (for details and exact definition see section \ref{sec:morphology}).

\autoref{fig:median_morphology} shows the median spin--filament alignment split according to the morphology of the central galaxy. Similar to \autoref{fig:median_angular_momentum}, we show the alignments of the halo, stellar and gaseous components as a function of both halo and stellar mass. \autoref{fig:median_morphology} shows a clear variation of the alignment signal with galaxy morphology.
This trend is the largest for the left-hand panel, indicating that host haloes of spheroids tend to have their spins more perpendicular to the filament axis than equal mass host haloes of disc galaxies. The same trend is also seen in the stellar/gas spin-filament alignment, although the trend is not as substantial as for the host haloes. Thus, in EAGLE, elliptical galaxies show a propensity towards a stronger orthogonal alignment than spirals, in qualitative agreement with observations. 

Although this result 
indicates that the halo spin orientation 
with respect to the large-scale structure 
affects the galaxy morphology, 
we found no significant evidence for this hypothesis. The fraction of spheroid central galaxies is roughly the same independent of the host halo spin--filament alignment. 
Thus, a more likely explanation for the variation of the halo spin--filament alignment with galaxy morphology is that they are both affected by a third physical process. For example, \citet{Welker2014} have shown that galaxy mergers, which typically take place along the filament in which the galaxies are embedded, can lead to an increase both in the fraction of spheroids and in the fraction of haloes and galaxies with spins perpendicular to their filaments.

Our results qualitatively match observational trends, in the sense that spheroids have an excess of perpendicular spin--filament alignments compared to disc galaxies, however they do not do so quantitatively. For example, \citet{tempel2013} have found that in observations spiral galaxies show a small, but statistically significant, preference to have their spins parallel to their host filament axis. However, in EAGLE we find that at all masses the spins of disc galaxies are preferentially perpendicular to their filaments. The discrepancy could be due to the difference in the mass range and environment of spiral galaxies between the EAGLE simulation and observations.
Due to magnitude limits, most observational analyses focus on generally bright and massive spirals. However, due to the small box size, EAGLE contains only a small number of such high mass spirals. 
Generally, such massive spirals are residing in thick filaments and it has been shown that haloes populating thicker filaments are more likely to have their spin aligned along the filament than equal mass haloes in thinner filaments \citep{GVeena2018}. Due to the limited size of the EAGLE simulation, there are only a few massive filaments and most spiral galaxies are found in thin filaments, which could explain the systematic difference between the EAGLE results and observations. 
We note that there are additional differences between observations and our EAGLE results that could also add to the discrepancy. Such as the different definitions of galaxy morphology and also that observational results are based on the alignment between the minor axis of galaxies and not their spin per-se. 

\subsection{Galaxy shape alignments}
\label{sec:shape_alignment} 
Observationally, it is very difficult to determine the spin of galaxies and we can only infer their shapes. In general, disc galaxies have their spin well aligned with their minor axis, however spheroid galaxies can have their spin and minor axis highly misaligned. For example, our sample of EAGLE spheroids have a median misalignment angle of $45^\circ$. Furthermore, even some spiral galaxies, e.g. those with a dominant bulge component, can have some degree of misalignment between their spin and minor axis. 

In \autoref{fig:median_shape_filament} we show the alignment of halo/galaxy minor axis with the orientation of their host filaments. Similar to spheroid galaxies, haloes are mostly dispersion supported and can have a large degree of misalignment between their spin and minor axis \cite[e.g. see][]{bett2007}. The left-hand panel of \autoref{fig:median_shape_filament} shows that the halo minor axis is preferentially perpendicular to the filament axis for objects of all masses. This is in contrast to the halo spin which shows a transition from preferentially perpendicular at high mass to preferentially parallel at low masses (see \autoref{fig:median_morphology}). The halo minor axis--filament alignment is the largest for haloes hosting spheroid galaxies and shows a mass dependence, being largest for high mass haloes (notwithstanding the highest mass bin which is affected by poor statistics due to the low number of objects). This is in agreement with the results of DM-only simulations, which also find that the halo shape--filament alignment is the largest for high mass haloes \citep[e.g. see][]{aragon2007,hahn2007,GVeena2018}.

The central panel of \autoref{fig:median_shape_filament} shows that the galaxy minor axis--filament alignment is different from the galaxy spin--filament alignment. For example, within our limited statistics we find that the galaxy minor axis--filament alignment in EAGLE is independent of galaxy mass. Furthermore, while spheroid galaxies have their minor axis preferentially perpendicular to the filament axis, spiral galaxies show no preferential alignment, that is their median alignment angle is consistent with $60^\circ$, which is the expectation for the no alignment case. The largest difference between spin and shape alignments of spiral galaxies is at the high mass end, $M_{\rm star}\geq5\times10^{10}$ \massUnit{}, where spirals have their spin preferentially perpendicular to the filament spine (see \autoref{fig:median_morphology}) but show no significant alignment in terms of their minor axis. The discrepancy could be due to many massive spirals having a significant bulge component whose spin, at least in EAGLE, is not within the plane of disc.

In summary, due to the degree of misalignment between galaxy spin and shape, the galaxy spin and galaxy minor axis show different alignments with their host filaments. This needs to be accounted for when comparing against observations, which can only measure galaxy shapes. Furthermore, we note that 
in most cases the 3D orientation of a galaxy is inferred from its projected 2D image and this can, in turn, affect the alignment signal; however we
leave this for further study. A more firm determination would be possible from velocity field maps, which can be obtained from 21cm radio observations
or from integral field spectrographs. While new generation instruments like the VLT MUSE facility and the wide-field APERTIF array on the WSRT
radio interferometer will be powerful resources, as yet the amount of available data for such large-scale alignment studies
is still rather limited. 

\begin{figure}
    \includegraphics[width=\columnwidth]{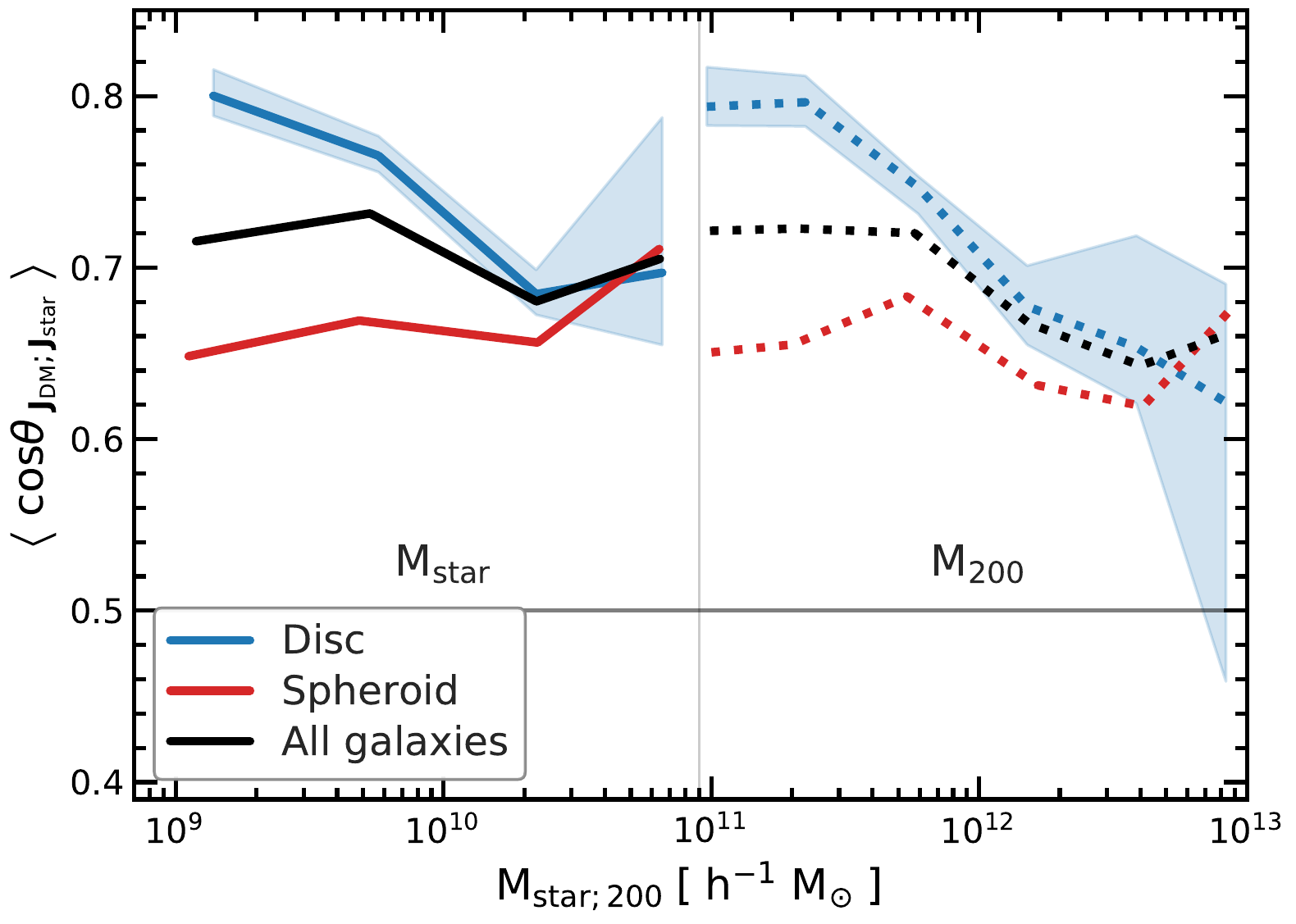}
    \includegraphics[width=\columnwidth]{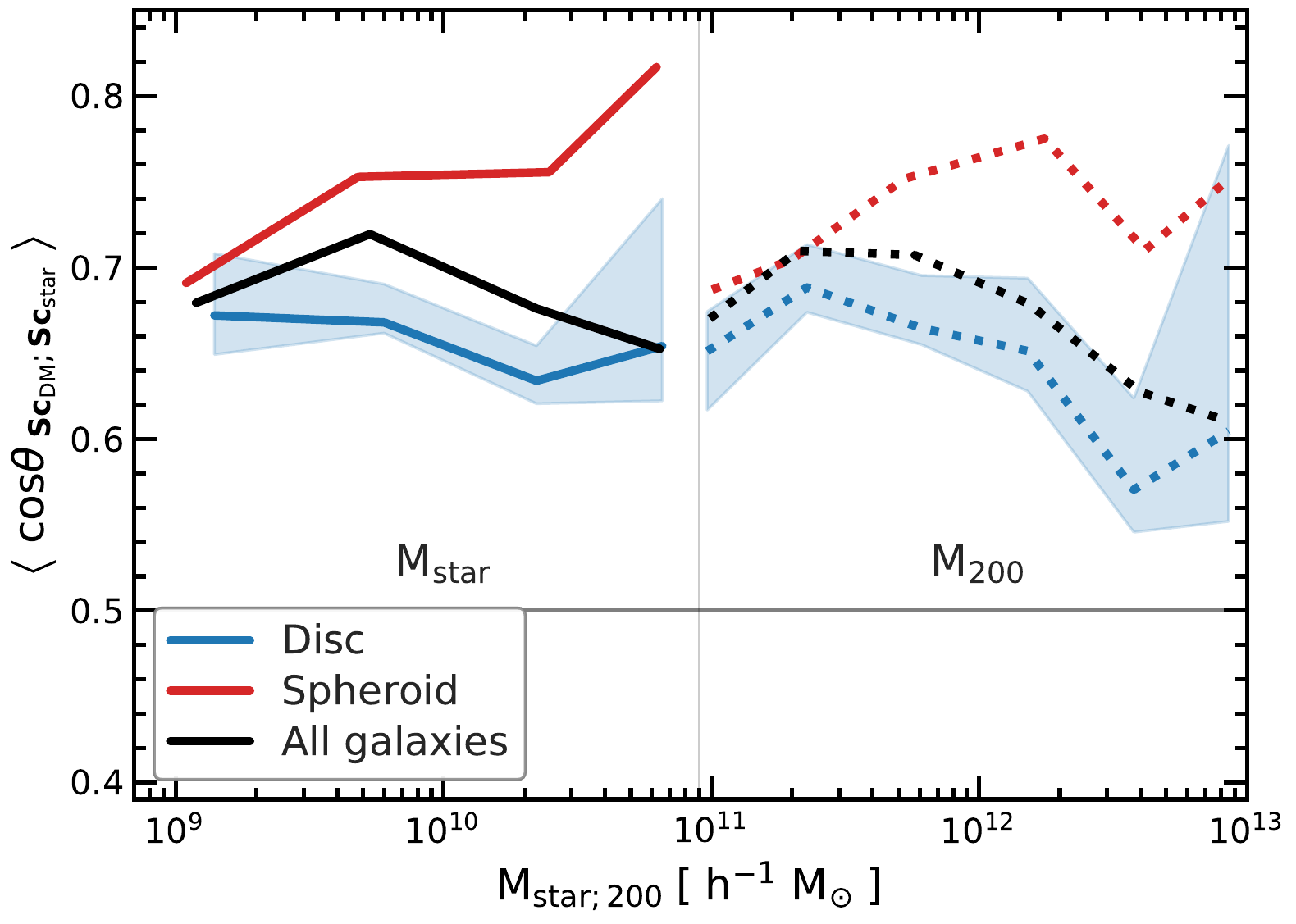}
    \vskip -.2cm
    \caption{\textbf{Galaxy--halo spin and shape alignment.} The plot shows the median alignment angle of the spin (top panel) and of the minor axis (bottom panel) between haloes and their central galaxies. The alignment is show as a function of both halo and stellar mass corresponds to all galaxies (black) as well as to two morphology selected subsamples: discs (blue) and spheroids (red).}
    \label{fig:median_galaxy_halo}
\end{figure}

\subsection{The halo--galaxy connection}
\label{sec:halo_galaxy} 
Galaxies and their host haloes form within the same large-scale environment, however, as we have shown in this section, haloes and galaxies are characterized by different spin--filament alignment trends. As discussed in the introduction, these differences are mostly due to the complex gas inflow and outflow physics which drives the formation of the stellar and gaseous components of a galaxy. Here, we study in more detail the spin and shape alignment of galaxies and their haloes, and how it relates to their host filament.

\autoref{fig:median_galaxy_halo} shows the spin and shape alignment between central galaxies and their host haloes. These are plotted as a function of both galaxy and halo mass and are also split according to the morphology of the central galaxy. In general, the galaxy spin shows a $45^\circ$ median misalignment angle with respect to the host halo spin (see top panel of \autoref{fig:median_galaxy_halo}). At low galaxy masses, this misalignment angle varies with galaxy morphology, with a median misalignment angle of $37^\circ$ and $49^\circ$ for disc and spheroids, respectively. 

Even though the spin of spheroid galaxies is the least well aligned with that of their host haloes, the spin--filament alignment of spheroid galaxies traces very well the spin--filament alignment of their host haloes (compare red curves between the left-hand and central panels of \autoref{fig:median_morphology}). Disc galaxies, in contrast, have a different spin--filament alignment than that of their host haloes: the spirals have a larger tendency to have their spins perpendicular to the filament axis than their host haloes. This difference is likely due to the dichotomy in the formation of disc and spheroid galaxies \citep[e.g.][]{Sales2012,Rodriguez-Gomez2017,Clauwens2018,Lagos2018}. The growth of spirals is thought to occur mostly through the accretion of gas with a coherently aligned angular momentum over a long period of time. Most spirals experience only uneventful minor mergers and their disc orientations vary slowly in time \citep[although there are exceptions. e.g. see][]{bett2012,Bett2016}. Such mergers are unlikely to misalign the spins of the central galaxy and the host halo. In contrast, a significant fraction of spheroids forms through major mergers, with the merger taking place preferentially along the filament
in which the galaxies are found \citep[e.g.][]{libeskind2014,Welker2014,Shao2018}. In this case, the mergers would preferentially orient the spin of both the galaxy and the halo perpendicular to the filament axis. 

The minor axis of galaxies and their host haloes shows a modest degree of alignment, with a median misalignment angle of $47^\circ$ that is roughly galaxy and halo mass independent, as can be appreciated from the bottom panel of \autoref{fig:median_galaxy_halo}. When splitting the sample according to galaxy morphology, we find a better alignment for spheroids than for spirals. This difference is the largest at the high mass end. The variation with galaxy morphology is opposite to the one found for galaxy--halo spin alignment, which shows a higher degree of alignment for disc galaxies. 

\subsubsection{Halo-galaxy alignments: the filament connection}
\label{sec:halo_galaxy_filament} 
The misalignment between galaxy and halo shapes explains why galaxy shapes are more poorly aligned with the filament axis than their host haloes. This is true for both disc and spheroid galaxies. In particular, the shape of haloes is aligned with their host filaments since it is mostly determined by recent accretion that takes place preferentially along the filament in which the halo is currently embedded \citep[see e.g.][and discussion therein]{GVeena2018}. In contrast, the shape of galaxies should be best aligned with the filament orientation when the galaxies formed most of their stellar mass, which took place at a redshift, $z\sim1-2$. The orientation of the host filament can change over time, due to either filament mergers or galaxies moving across the cosmic web \citep{haarlem1993,aragon2007,cautun2014,Wang2017a}, and thus should decrease in time.

\begin{figure}
    \includegraphics[width=\columnwidth]{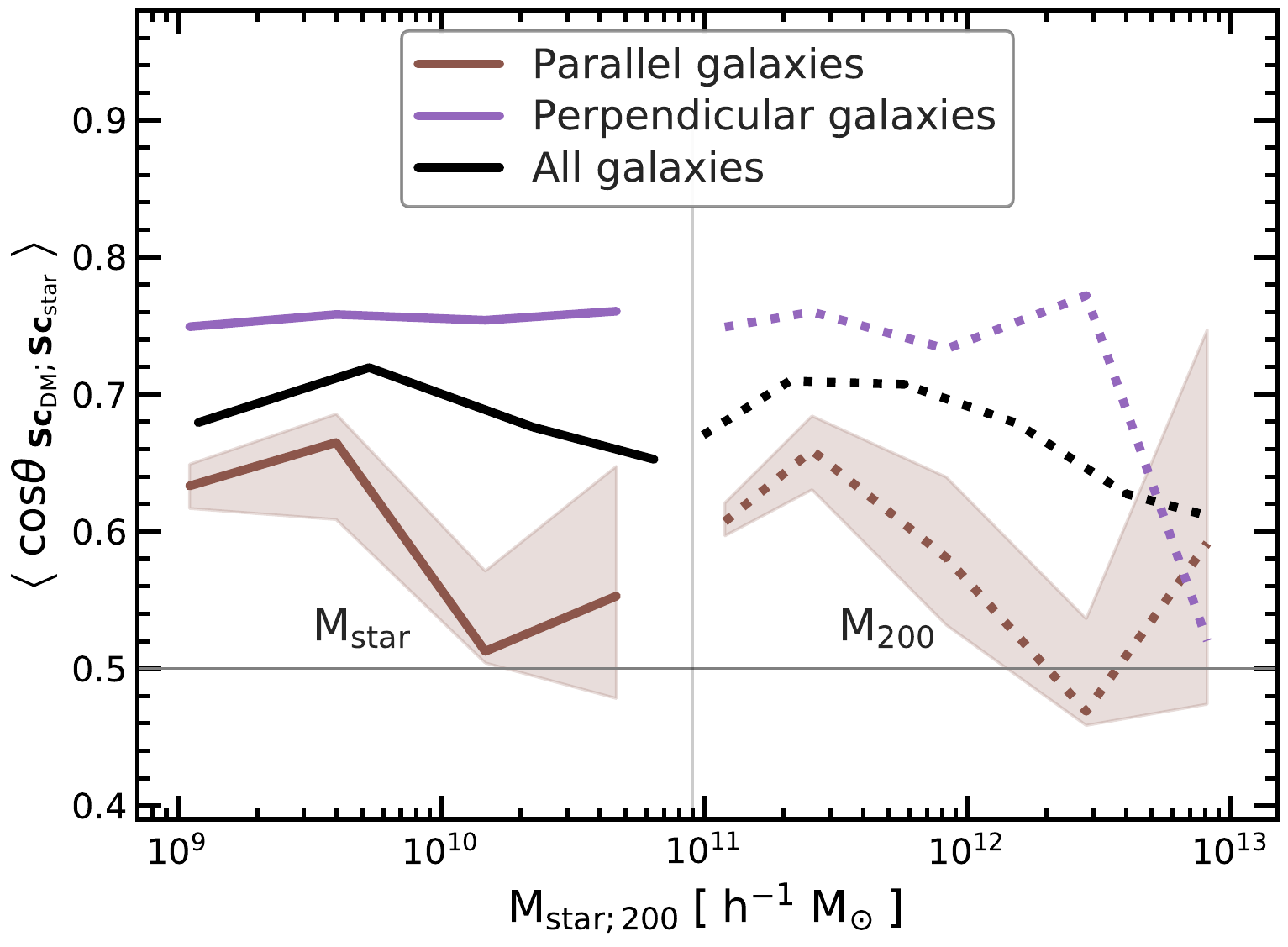}
    \vskip -.2cm
    \caption{\textbf{Galaxy--halo minor axis alignment for galaxy subsamples selected according to the galaxy minor axis--filament alignment angle.} 
    The plot shows the median alignment angle between the minor axis of central galaxies and that of their host haloes. We show results for all galaxies (black) as well as for two galaxy samples selected to have their minor axis along the filament (so called parallel galaxies shown in purple) or perpendicular to the filament (so called perpendicular galaxies shown in brown). Galaxies perpendicular to filaments show a larger alignment with their host haloes.}
    \label{fig:median_galaxy_populations}
\end{figure}

To better understand the processes affecting the galaxy--filament alignment, we proceed by selecting two galaxy subsamples: one composed of galaxies that have their minor axis parallel to the filament and a second one composed of galaxies with minor axis perpendicular to the filament. For clarity, we refer to the two subsamples as parallel and perpendicular galaxies. To have a large enough sample, we define the galaxies with parallel minor axis--filament orientations as those with small misalignment angles, that is those with $\cos \theta_{{\rm \mathbf{Sc}_{star}}, \mathbf{e}_3} \geq 0.8$. Similarly, we define the galaxies with perpendicular minor axis--filament orientations as those with a misalignment angle close to $90^\circ$, that is those with $\cos \theta_{{\rm \mathbf{Sc}_{star}}, \mathbf{e}_3} \leq 0.2$. Each of the two subsamples consist of roughly $20\%$ of the total galaxy population. 

Galaxy and halo shapes are moderately aligned (see e.g. \autoref{fig:median_galaxy_halo}) and thus it should not be surprising that parallel galaxies reside in haloes whose minor axis is also predominately parallel to the filament axis. Similarly, perpendicular galaxies are found in haloes whose minor axis is predominately perpendicular to the filament axis. 
More interestingly is to study how the galaxy--halo shape alignment varies between perpendicular and parallel galaxies. This is because the galaxy minor axis--filament alignment is weak and should not affect noticeably the galaxy--halo alignment.

\autoref{fig:median_galaxy_populations} shows the median galaxy--halo minor axis alignment angle for the two subsamples of parallel and perpendicular galaxies. It clearly highlights that galaxies perpendicular to their filaments have a larger degree of alignment with their host haloes. In contrast, galaxies oriented along their filament axis have poorer alignments with their haloes. It suggest that the same processes that affect the galaxy shape--filament alignment play an important role for the galaxy--halo alignment too. For examples, galaxies and haloes embedded in filaments that remain stable over long periods of time are more likely to experience anisotropic infall along the same time-independent directions. This would lead to a stronger alignment between the galaxy and its halo \citep{haarlem1993}. Furthermore, this would also lead to a preferentially perpendicular alignment of galaxy and halo minor axes with the host filament since accretion preferentially takes place along the filament direction. On the other hand, objects whose cosmic web environment changes rapidly with time experience different anisotropic infall directions at various times. Most of the galaxy stellar mass is acquired at early times, while haloes are still assembling at late times. Thus, on average, such galaxies are more poorly aligned with their haloes.

A similar dichotomy in galaxy--halo alignment is present when selecting parallel and perpendicular galaxies subsamples according to the galaxy spin--filament alignment. In this case, we also find that galaxies with spins perpendicular to their filaments are better aligned with their host haloes.
While the effect is about half the size of the one seen in \autoref{fig:median_galaxy_populations} and given the similarity, for brevity we do not
include a diagram to illustrate it.

\subsubsection{Implications: satellite planes, halo shapes}
\label{sec:halo_galaxy_implications} 
The results illustrated in \autoref{fig:median_galaxy_populations}, which are that galaxies whose minor axis is perpendicular to their host filament are more likely to be aligned with their host haloes, have two important implications. \citet{Shao2016} have found a similar result when studying the alignment between central galaxies and their satellite galaxies: systems where most satellites are in the plane of the baryonic disc have a much higher galaxy--halo alignment. Thus, combining our results and those of \citeauthor{Shao2016}, we predict that galaxies perpendicular to filaments should have most of their satellites in the plane of the galaxy disc. This prediction of the EAGLE simulation can be checked observationally and represents one avenue for constraining the processes that affect the alignment of galaxies and haloes with their host filaments.

Furthermore, our results can be used to test a fundamental prediction of the standard cosmological model: that galaxies are embedded in flattened dark matter haloes. A possible test of this prediction would involve stacking weak lensing maps of multiple galaxies and measuring the average flattening of their DM halo \citep[see e.g.][]{van_Uitert2017}. To do so, one needs to know the orientation of the DM halo. Using galaxy orientations does not work due to galaxy--halo misalignment and makes it very challenging to measure halo shapes \citep[see e.g.][and references therein]{Bett2012b}. Selecting a subset of galaxies perpendicular on their filament improves the galaxy--halo alignment and could represent an improved approach for measuring halo shapes.

\section{Conclusions and Discussions}
The principal purpose of this study is to investigate how far secondary processes alter the original spin of haloes and galaxies. While the spin of haloes and galaxies is initially generated by tidal torques \citep{hoyle1949, peebles1969, doroshkevich1970,white1984}, a range
of nonlinear and baryonic processes are likely to alter the evolution of this fundamental property of galaxies. Because the large-scale tidal field is
the agent behind the contraction of mass into elongated filaments, the tidal torque theory leads to the expectation that halo and galaxy spins should tend to be oriented perpendicular to the filaments in which they are embedded \citep[see e.g.][]{lee2000,joneswey2009}. Hence, by relating the
spin of haloes and galaxies to their cosmic web environment we seek to identify the processes which affect the rotation of haloes and galaxies. 

The simulation-based studies of \citet{aragon2007} and \citet{hahn2007} were the first to reveal that the halo spin--filament alignment shows a complex mass dependence, with high mass
haloes having their spins preferentially perpendicular to filaments while low mass haloes show the opposite trend. The transition between the two regimes today takes place at a halo mass of ${\sim}10^{12}$ \massUnit{}, which is known as the spin-flip transition mass. In \citet{GVeena2018} we performed a detailed and systematic analysis of halo spin--filament alignments to reveal a strong dependence of the alignment on the nature of filaments. In particular, the spin-flip transition mass is highest for haloes in dynamically dominant and thick filaments, while it is an order of magnitude lower for haloes in thin and tenuous
filaments. This trend represents a clear indication of the impact of late-time halo evolution processes on the spin--filament alignment.

The present study extends our investigation from DM haloes to the galaxies they harbour. For this, we employ the EAGLE hydrodynamical simulation, which follows the formation and evolution of galaxies in a cosmological volume. Our goal is to address the question in how far the alignment of haloes with respect to the filaments in which they reside is reflected
in an alignment of galaxies with respect to the large scale cosmic web. To this end, we investigate whether the mass dependent alignment
exhibited by haloes is also exhibited by galaxies. In particular, we wish to assess the relation
between the spin orientations of galaxies and their host haloes, and establish to what extent the halo spin orientation can be inferred from that of their central galaxies. In this context, we also investigate if the presence of baryons alters the halo spin--filament alignment found in DM-only simulations. 

\subsection*{Filament population}
We predominantly focus on cosmic filaments. We have identified the filament population with two different cosmic web finders
\citep[see e.g.][]{Libeskind2018} in order to assess which results are dependent on the filament identification method. 
Firstly, we applied the NEXUS+ multiscale formalism to the matter distribution to identify the cosmic web nodes, filaments, sheets and voids. Secondly, we applied the Bisous method to the galaxy distribution to select galaxy filaments. In general, we find good agreement between the filament populations identified by the two methods, although some differences exist (see \autoref{sec:difference_fila_population} for a detailed analysis). In particular, we find a large overlap between the populations of filament galaxies in the two methods and good alignment between the orientations of the NEXUS+ and Bisous filaments. These are the two aspects that are most important for this study.

We find that most of the matter content of the $z=0$ universe is found in filaments, in good agreement with cosmic web studies based on DM-only simulations \citep[e.g.][]{cautun2014,Libeskind2018}. More specifically, 52\% of the DM is found in filaments, but only 47\% of the total gas content.
The difference is due to baryonic processes that heat up the gas in filaments and disperse it to surrounding walls and voids \citep{haider2016}. Accordingly, we find that walls and voids contain slightly higher mass fractions of gas than of DM. 

The majority of the stellar mass -- 82\% for NEXUS+ and 70\% for Bisous -- is located in filaments. The remaining stellar mass is found in nodes and sheets, each containing roughly 10\% of the stars, while voids contain less than 0.5\% of the total stellar mass. The filaments also contain the largest number of galaxies with stellar masses higher than $10^9$ \massUnit{}, and a dominant fraction of lower mass galaxies. In particular, most of the galaxies at the knee of the stellar mass function are found in filaments, while the very massive galaxies are found mostly in the cosmic web nodes. On the other hand, voids are mostly populated by faint dwarf galaxies.

\subsection*{Halo and galaxy alignments with filaments}
We have investigated four different, but intimately related, aspect of the alignment between galaxies, haloes and filaments. These are the alignments of
\begin{itemize}
    \item[$\bullet$] the spin of haloes and the spin of galaxies with respect to the embedding filament, 
    \item[$\bullet$] the spin of late-type and early-type galaxies relative to the filament direction,
    \item[$\bullet$] the shape of haloes and galaxies, characterised by their minor axis, with respect to the filament in which they reside, and
    \item[$\bullet$] the spins of haloes with respect to the galaxies they host, as well as the orientation of the minor axis of galaxies with respect to that of their haloes. 
\end{itemize}

\subsubsection*{Halo and galaxy spin alignment with filaments}
We find that haloes in the DM-only P-Millennium simulation and the EAGLE
hydrodynamical simulation show statistically similar distributions of spin--filament alignments: low mass haloes have their spins preferentially along the filament while high mass haloes have their spin preferentially perpendicular to the filament. The halo transition mass between the two regimes is
the same for both DM-only and baryonic physics simulations. Thus, the addition of baryons does not affect the distribution of halo spins with respect
to the large-scale filaments. 

Galaxies, just as their host haloes, show a mass dependent spin--filament alignment: massive galaxies have their spin preferentially perpendicular to the filament to a larger extent than lower mass objects. However, we do not detect a clear transition from parallel to perpendicular alignment as we see in the case of haloes. This is in contrast to the results of \citet{Wang2018} who, using the Illustris simulation, found a transition in the galaxy spin--filament alignment that takes place at a stellar mass of $2.5 \times 10^{9}$ \massUnit{}. However, our results agree better with the  \citet{Codis2018} study based on the Horizon-AGN simulation, which shows that $z=0$ galaxies with stellar masses below $10^{10}$ \massUnit{} have no preferential spin alignment with their host filament. 

The discrepancy with the studies of \cite{Wang2018} and \cite{Codis2018} could be due to difference in hydrodynamic simulations.
In particular, many of the subgrid implementations of baryonic physics vary from simulation to simulation and this can lead to different
galaxy growth histories and thus to different galaxy spins orientations. However, we suspect that at least part of the discrepancy is a manifestation
of the dependence of the galaxy spin--filament alignment on the nature of filaments. For example, this trend has been robustly established by
\citet{GVeena2018} for the halo spin--filament alignment, with equal mass haloes having a larger propensity for perpendicular spin--filament alignments
when they are found in thinner filaments. This trend is strong enough to result in more than one order of magnitude variation in the halo spin-flip
mass between the thinnest and thickest filaments.
We also find tentative evidence for a dependence between the alignment of galaxy spins and filament properties; however the EAGLE volume is too small to robustly characterise such a trend.

We note that a systematic trend in which galaxies in thinner filaments are more likely to have their spins perpendicular to the filament axis than
similar mass galaxies in thicker filaments is consistent with the differences between our alignment results
and those of  \citet{Wang2018} and \citet{Codis2018}. The \citeauthor{Wang2018} method identifies mostly thick filaments, while both the NEXUS+ and
Bisous algorithms detect a large population of very thin filaments. The DISPERSE method employed by \citeauthor{Codis2018} is somewhere in between the
other two studies \citep[for a detailed comparison, see][]{Libeskind2018}.

\subsubsection*{Dependence on galaxy morphology}
We find a strong dependence of the halo and galaxy spin--filament alignment on galaxy morphology. Distinguishing between disc and spheroid
galaxy populations, we find that the host haloes of spheroidal galaxies show a larger tendency to have perpendicular spin--filament alignments than the host haloes of disc galaxies. Similarly, spheroid galaxies show a stronger propensity for an orthogonal spin-filament alignment than spirals.
This agrees with observations which find that ellipticals are more likely to be orthogonal to filaments than spirals
\citep{tempel2013,Zhang2013,Zhang2015,Pahwa2016}, although we do not detect a distinct parallel alignment of disc galaxies as detected in the SDSS. 

The discrepancy could be due to the small volume of the EAGLE simulation. Due to a lack of very large modes, the simulation does not contain
many objects similar to those typically present in observations: bright spiral galaxies as well as the thick filaments which host them.

\subsubsection*{Shape alignment with filaments}
We find that many galaxies show a misalignment between their minor axis and their spin. The misalignment is largest for elliptical galaxies.
However, even disc-dominated objects which have a massive bulge fraction can show some degree of misalignment. This results in a difference between
the alignments of galaxy spin and minor axis with the filament.

Interestingly, we do find that the galaxy minor axis--filament alignment is largely mass independent. We also find that ellipticals show a larger
degree of orthogonal alignment than spirals. On the other hand, there is no significant evidence for a preferential alignment of the minor
axis of spiral galaxies with the spine of filaments.

\subsubsection*{Perpendicular versus parallel filament galaxies}
To study the processes responsible for the galaxy--filament alignment, we split or sample according to the galaxy minor axis--filament misalignment
angle. We have selected the 20\% of the population with the closest to perpendicular orientations as well as an equal fraction with the closest to parallel
orientations. We find that galaxies with a perpendicular orientation relative to their filament are much better aligned with their host halo than
the population as a whole. In contrast, galaxies parallel to their filament are poorly aligned with their halo.

This suggests that the same processes that affect the galaxy--filament alignment are at least partially responsible for the galaxy--halo alignment too.
One such factor could be the coherence over long periods of time of the cosmic web around a halo/galaxy. Objects embedded in such filaments
experience anisotropic infall along the same time-independent directions \citep{haarlem1993}. Such objects are expected to have a
better alignment between haloes and the galaxies they harbour, as well as with their host filaments. On the other hand, in a rapidly
changing environment objects experience anisotropic infall directions that vary with time, implying a higher degree of misalignment
\citep[see e.g.][]{haarlem1993}.

\subsection*{Summary}
To summarise, we have studied the present day alignments of the shapes and spins of haloes and galaxies with their host filament in the EAGLE simulation. This
represents a first step towards understanding the processes that determine these alignments with the large-scale cosmic web.

The alignments we have studied are weak and to properly characterize them we need a large sample of galaxies. This is difficult with current
hydrodynamics simulations, which typically follow relatively small cosmological volumes. Thus, it is critical to expand the study to larger simulations,
such as Illustris300 which has a 27 times larger volume than EAGLE \citep{Pillepich2018}, and to robustly quantify secondary trends, such as the dependence of the
alignment signal on filament properties. In parallel with this approach, we also need to understand how the alignments vary in time. In particular,
it is critical to follow the evolution of individual haloes and galaxies and understand what factors affect their alignment with the filament axis.
This is the subject of the third, upcoming, paper in our study. 

\section*{Acknowledgements}
We are grateful to Bernard Jones, Enn Saar and Noam Libeskind for many encouraging and insightful discussions, several of which formed the
inspiration for the present study. PGV thanks the Institute of Computational Cosmology (ICC) in Durham and the 
Leibniz Institute for Astrophysics in Potsdam for their support and hospitality during several long work visits during which a major share of
the work for this study was carried out. Also, RvdW thanks the ICC for its hospitality and support during several short work visits. MC and CSF
were supported by the Science and Technology Facilities Council (STFC) [grant number ST/I00162X/1, ST/P000541/1] and by the ERC Advanced Investigator
grant, DMIDAS [GA 786910]. ET acknowledges the support of the ETAg grants IUT26-2, IUT40-2, and of the European Regional Development Fund (TK133, MOBTP86).
This work used the DiRAC@Durham facility managed by the Institute for Computational Cosmology on behalf of the STFC DiRAC HPC Facility (www.dirac.ac.uk). The equipment was funded by BEIS capital funding via STFC capital grants ST/K00042X/1, ST/P002293/1, ST/R002371/1 and ST/S002502/1, Durham University and STFC operations grant ST/R000832/1. DiRAC is part of the National e-Infrastructure.

\bibliographystyle{mnras}
\bibliography{bibliography}





\bsp	
\label{lastpage}

\end{document}